\documentclass[tighten,preprint2,flushrt]{aastex}
\usepackage{geometry}
\usepackage{float}
\usepackage{slashbox}
\usepackage{multirow}
\usepackage[hyphens]{url}
\usepackage{subfigure}
\usepackage{color}
\usepackage{color,soul}
\usepackage{rotating}
\newcommand{\myemail}{zouhu@nao.cas.cn}
\newcommand{\boldtext}[1]{\textbf{\boldmath #1}}

\begin{document}

\title{The First Data Release of the Beijing-Arizona Sky Survey}
\author{Hu Zou\altaffilmark{1}, Tianmeng Zhang\altaffilmark{1,2}, Zhimin Zhou\altaffilmark{1}, Jundan Nie\altaffilmark{1}, Xiyan Peng\altaffilmark{1}, Xu Zhou\altaffilmark{1}, Linhua Jiang\altaffilmark{3}, Zheng Cai\altaffilmark{4}, Arjun Dey\altaffilmark{5}, Xiaohui Fan\altaffilmark{4}, Dongwei Fan\altaffilmark{1}, Yucheng Guo\altaffilmark{3,6}, Boliang He\altaffilmark{1}, Zhaoji Jiang\altaffilmark{1},  Dustin Lang\altaffilmark{7}, Michael Lesser \altaffilmark{4}, Zefeng Li\altaffilmark{4}, Jun Ma\altaffilmark{1,2}, Shude Mao\altaffilmark{8,1,9}, Ian McGreer\altaffilmark{4}, David Schlegel\altaffilmark{10}, Yali Shao\altaffilmark{3,6}, Jiali Wang\altaffilmark{1}, Shu Wang\altaffilmark{3,6}, Jin Wu\altaffilmark{3,6}, Xiaohan Wu\altaffilmark{4}, Qian Yang\altaffilmark{4}, Minghao Yue\altaffilmark{3,6}} 
\altaffiltext{1}{Key Laboratory of Optical Astronomy, National Astronomical Observatories, Chinese Academy of Sciences, Beijing 100012, China; \myemail}
\altaffiltext{2}{College of Astronomy and Space Sciences, University of Chinese Academy of Sciences, Beijing 100049, China}
\altaffiltext{3}{Kavli Institute for Astronomy and Astrophysics, Peking University, Beijing 100871, China}
\altaffiltext{4}{Steward Observatory, University of Arizona, Tucson, AZ 85721}
\altaffiltext{5}{National Optical Astronomy Observatory, Tucson, AZ 85719}
\altaffiltext{6}{Department of Astronomy, School of Physics, Peking University, Beijing 100871, China}
\altaffiltext{7}{David Dunlap Institute, University of Toronto, Toronto, Canada}
\altaffiltext{8}{Department of Physics and Tsinghua Center for Astrophysics, Tsinghua University, Beijing 100086, China}
\altaffiltext{9}{Jodrell Bank Centre for Astrophysics, University of Manchester, Manchester M13 9PL, UK}
\altaffiltext{10}{Lawrence Berkeley National Labortatory, Berkeley, CA 94720}

\begin{abstract} 
The Beijing-Arizona Sky Survey (BASS) is a new wide-field legacy imaging survey in the northern Galactic cap using the 2.3m Bok telescope.  The survey will cover about 5400 deg$^2$ in the $g$ and $r$ bands, and the expected 5$\sigma$ depths (corrected for the Galactic extinction) in these two bands are \boldtext{$g = 24.0$ and $r = 23.4$ mag (AB magnitude)}.  BASS started observations in January 2015, and has completed about 41\% of the whole area as of July 2016. The first data release contains calibrated images obtained in 2015 and 2016 and their corresponding single-epoch catalogs and co-added catalogs . The actual depths of single-epoch images are \boldtext{$g \sim 23.4$ and $r \sim 22.9$ mag}. The full depths of three epochs are \boldtext{$g \sim 24.1$ and $r \sim 23.5$ mag}.
\end{abstract}


\section{Introduction}
BASS \footnote{\url{http://batc.bao.ac.cn/BASS/}} is a new $g$ and $r$-band imaging survey conducted by the National Astronomical Observatory of China (NAOC) and Steward Observatory \citep{zou17}, following the success of the South Galactic Cap $u$-band Sky Survey \citep[SCUSS;][]{zho16}. Both surveys use the  2.3 m Bok telescope located on Kitt Peak and its 90Prime imager (built with the funding support from the SCUSS project). BASS will survey about 5400 deg$^2$ in the north Galactic cap. The 5$\sigma$ depths when corrected for the Galactic extinction are expected to be 24.0 and 23.4 mag in the $g$ and $r$ bands, respectively. BASS is also one of three optical imaging surveys that are designated for target selection of the Dark Energy Spectroscopic Instrument \citep[DESI;][]{des16} project. The other two surveys are the Dark Energy Camera Legacy Survey \citep[DECaLS;][]{blu16} and the MOSAIC $z$-band Legacy Survey \citep[MzLS;][]{sil16}. The DESI spectroscopic survey will obtain more than 30 million spectra, an order of magnitude larger than the Sloan Digital Sky Survey \citep[SDSS;][]{yor00}. It will have a profound impact on studies of the Milk Way, galaxies, and cosmology. BASS, together with the other two DESI imaging surveys, will cover a total area of 14,000 deg$^2$, and will provide unique science opportunities in studying the Galactic halo substructures,  satellite dwarf galaxies around the Milk Way, galaxy clusters, high-redshift quasars, and so on \citep[see scientific descriptions in][]{zou17}.

The BASS observations started in 2015. Much effort has been made to improve the telescope control system, instrument, and ancillary observing tools such as the exposure time calculator and real-time displays of image qualities, weather conditions, and fields to be observed at night on the figure of the all-sky camera \citep[see more details in][]{zou17}. We have completed about 41\% of the whole survey as of July 2016. In order to meet the BASS/DESI scientific requirements, we have developed new data reduction pipelines based on the SCUSS pipelines. This paper aims to describe the details of the first data release (DR1). Section \ref{sec-survey} gives a summary of the survey, including the instrument and the survey footprint. Section \ref{sec-observation} presents the survey progress and image quality. The detailed data reduction including the imaging processing, astrometry, and photometric calibration are described in Section \ref{sec-reduction}. The method of photometry is presented in Section \ref{sec-photometry}. Section \ref{sec-analyses} shows some analyses with the catalogs. Section \ref{sec-access} describes the data products and ways to access the data. The summary is given in Section \ref{sec-summary}.

\section{The Survey Instrument and Footprint} \label{sec-survey}
\subsection{Camera}
\begin{figure}[!ht]
\centering
\includegraphics[width=\columnwidth]{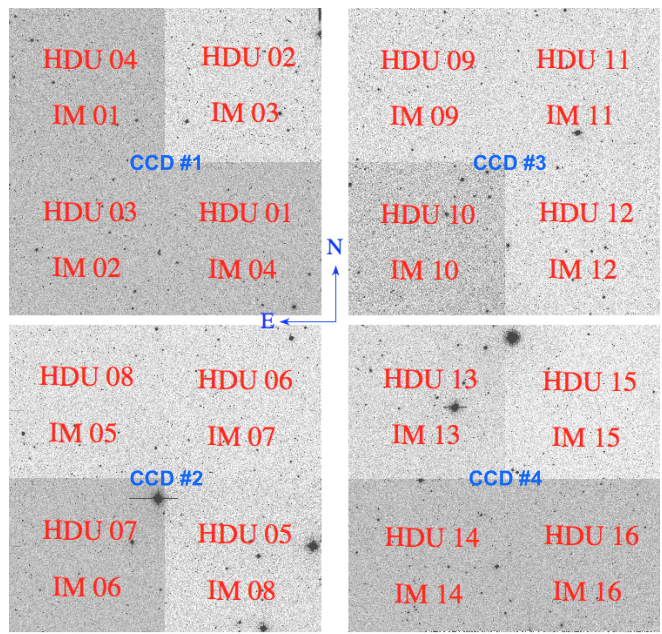}
\caption{Layout of the CCD mosaic. North is up and east to the left.  The blue labels represent the CCD numbers. The labels starting with ``IM" show the extinction names in the FITS headers, and those starting with ``HDU" show the extension numbers. The gaps in the north-south and east-west directions are 2\arcmin.8 and 0\arcmin.9, respectively.} \label{fig-ccdlayout}
\end{figure}

Table \ref{tab-summary} summarizses the BASS survey, including the telescope, camera, filter, and survey parameters. The Steward Observatory Bok telescope is a 2.3 m equatorial mounting telescope (see more details on its website\footnote{\url{http://james.as.arizona.edu/~psmith/90inch/90inch.html}}).  The 90Prime camera is installed at the prime focus of the telescope. It consists of four 4k by 4k back-illumination CCDs. The field of view (FoV) is about 1\arcdeg.08$\times$1\arcdeg.03 with inter-CCD gaps of 2\arcmin.8 along right ascension and 0\arcmin.9 along declination. The pixel scale is about 0\arcsec.454. These CCDs are optimized for the ultraviolet response. The quantum efficiencies in the $g$ and $r$ bands are higher than 90\% and 80\%, respectively. Each CCD has 4 amplifiers and thus there are 16 extensions in FITS images. Figure \ref{fig-ccdlayout} shows the layout of the CCD mosaic. If not specified, the layout of the mosaic images presented in this paper is the same as shown this figure. Table \ref{tab-gainrd} lists the gain and readout noise for each amplifier. The average values are 1.5 e/ADU and 8.4 e. The readout noise for CCDs \#2 and \#3 is relatively higher than that of the other two CCDs. 

\begin{table}[!ht]
\caption{The BASS survey summary}\label{tab-summary}
\small
  \begin{tabular}{rl}
  \tableline\tableline
  \multicolumn{2}{c}{Telescope}\\
  \tableline
  Telescope & 2.3m Bok telescope \\
  Corrected focal ratio & f/2.98 \\
  Absolute pointing & $<$3\arcsec RMS\\
  Site & Kitt Peak \\
  Elevation & 2071 m \\
    \tableline
   \multicolumn{2}{c}{Camera}\\
    \tableline
   CCD number & 4 \\
   CCD size & 4096$\times$4032 \\
   Pixel size & 0\arcsec.454 (15 $\mu$m) \\
   FoV & 1\arcdeg.08$\times$1\arcdeg.03 \\
   Full well & 90,000 e \\
   Gain & 1.5 e/ADU \\
   Readout noise & 8.4 e \\
   Readout time & 35 s \\
   \tableline
   \multicolumn{2}{c}{Filter}\\
   \tableline
   SDSS $g$ & 4776 \AA \\
   DES $r$ & 6412 \AA \\
   \tableline
   \multicolumn{2}{c}{Survey parameters}\\
   \tableline
   Area & 5400 deg$^2$ ($\delta > 30$\arcdeg) \\
   Depth (5$\sigma$) & $g = 24.0$, $r = 23.4$ \\
   Typical exposure time & 3$\times$100s \\
   Observation period & 2015--2018 \\
  \tableline
  \end{tabular}
\end{table}

\begin{table*}[!ht]
\caption{Gain and readout noise for each CCD}\label{tab-gainrd}
{\scriptsize
  \begin{tabular}{c|cccc|cccc|cccc|cccc}
  \tableline\tableline
CCD & \multicolumn{4}{c|}{CCD \#1} & \multicolumn{4}{c|}{CCD \#2} & \multicolumn{4}{c|}{CCD \#3} & \multicolumn{4}{c}{CCD \#4} \\
\tableline
HDU & 1 & 2 & 3 & 4 & 5 & 6 & 7 & 8 & 9 & 10 & 11 & 12 & 13 & 14 & 15 & 16\\
Gain &1.495 & 1.620 &1.524& 1.519 & 1.546& 1.656& 1.608& 1.590&1.399&1.503&1.425&1.553&1.505&1.535&1.521&1.584\\
Noise &7.353&7.288&7.313&6.597&9.299&9.509&9.302&8.006&10.122&8.897&9.474&9.019&9.527&7.940&6.903&7.716\\
 \tableline 
  \end{tabular}
}
\end{table*}

\subsection{Filters}
\begin{figure}[!ht]
\centering
\includegraphics[width=\columnwidth]{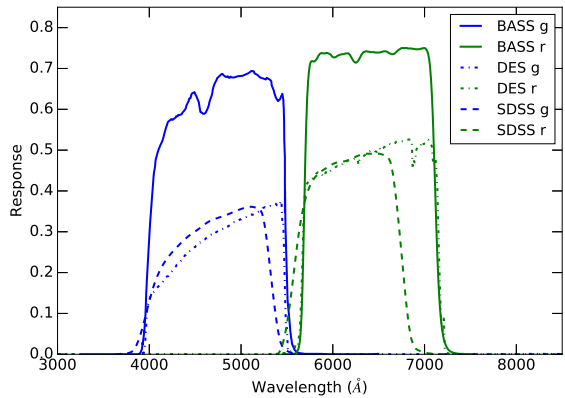}
\caption{Filter response curves for different surveys (solid for  BASS, dash-dotted for DES, and dashed for  SDSS). The $g$ and $r$-band responses are plotted in blue and green, respectively. } \label{fig-filter} 
\end{figure}

The BASS photometric system includes two broad bands ($g$ and $r$), which are consistent with those used in the Dark Energy Survey \citep[DES;][]{des05}. The other two DESI  imaging surveys (DECaLS and MzLS) also adopt the same filters. This helps us achieve homogeneous spectroscopic target selection.  The BASS $g$ filter is the existing SDSS $g$ filter that was used on the Bok telescope. This filter is fairly similar to the DES $g$ filter. The SDSS $r$ filter on the Bok is quite different from the DES one, so a new DES $r$ filter was purchased for BASS. Figure \ref{fig-filter} compares the filter response curves for different surveys. Table \ref{tab-filter} lists the effective wavelengths and bandwidths for these filters. The BASS filter response curves\footnote{\url{http://batc.bao.ac.cn/BASS/doku.php?id=datarelease:telescope_and_instrument:home#filters}} include the filter transmission, CCD quantum efficiency, and the atmospheric extinction at airmass of 1.0 \citep[see details in][]{zou17}. The SDSS filter responses also include the atmospheric extinction at airmass of 1.3 \citep{doi10}. The DES response is the total throughput\footnote{\url{http://www.ctio.noao.edu/noao/content/dark-energy-camera-decam}}. We can see that the SDSS $r$ band is bluer and narrower than the BASS/DES $r$ band. 

\begin{table*}[!ht]
\centering
\caption{Filter parameters for different surveys} \label{tab-filter} 
 \begin{tabular}{c|ccc|ccc}
\tableline\tableline
Filter & \multicolumn{3}{c|}{g} & \multicolumn{3}{c}{r} \\
\tableline
Survey & BASS & DES & SDSS & BASS & DES & SDSS \\
\tableline
Effective wavelength (\AA) & 4776 & 4842 & 4719 &  6412 & 6439 & 6185 \\
Bandwidth (\AA) & 848 &  966 & 777 & 833 & 895 & 687  \\
 \tableline 
 \end{tabular}
\end{table*}

\subsection{Footprint and tiling}
\begin{figure*}[!ht]
\centering
\includegraphics[width=\textwidth]{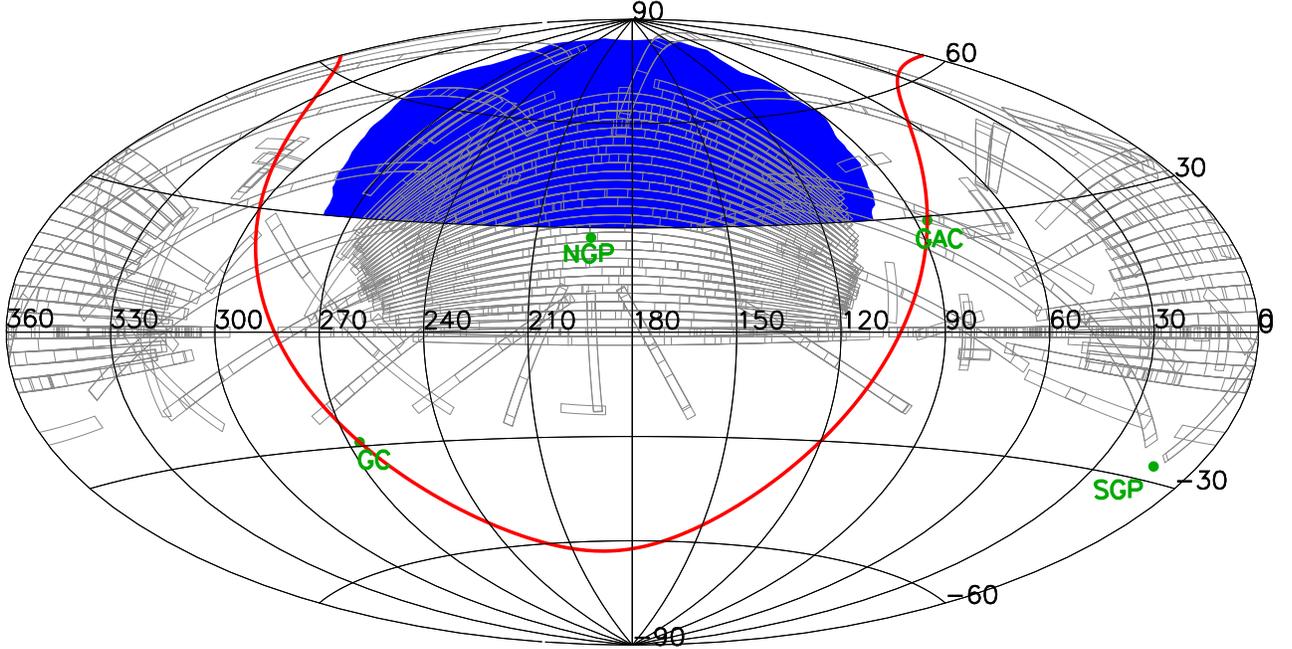}
\caption{\boldtext{BASS footprint (blue area) in the Aitoff projection of equatorial coordinate system. The projection is centered at ($\alpha$ = 180\arcdeg, $\delta$ = 0\arcdeg).} The SDSS imaging footprint are also plotted in grey. The green points are the Galactic center (GC), Galactic anticenter (GAC), north Galactic pole (NGP), and south Galactic pole (SGP). The red curve shows the Galactic plane.} \label{fig-footprint} 
\end{figure*}

The BASS coverage as shown in Figure \ref{fig-footprint} is a part of the DESI imaging footprint. It is located in the north of the northern Galactic cap. It is approximately enclosed by the declination of $\delta > 30\arcdeg$ and Galactic latitude of $b > 18\arcdeg$. The regions with large Galactic extinctions at high declinations are removed. Compared to the SDSS coverage, BASS will extend closer to the Galactic plane. The BASS footprint will be covered in three passes/epochs (or exposures). Each pass is divided into 5494 tiles that are evenly distributed over the footprint. The tile size is the same as the camera FoV, and adjacent tiles slightly overlap. The three passes are dithered by about 1/4 FoV. In this way, about 90.5\% of the area will be covered by all three passes or exposures, and a considerable fraction of CCD gaps are also covered by three exposures. These three exposures are taken under different weather conditions: Pass 1 for photometric nights and good seeing ($< 1\arcsec.7$); Pass 2 for either photometric nights or good seeing; Pass 3 for any other conditions \citep[also see in][]{zou17}. 

\section{Observation} \label{sec-observation}
\subsection{Survey progress}
BASS started its observations in January 2015 and has obtained 144 nights for 2015A and 2016A so far. Most of the nights in 2015 were grey or bright nights. In addition, the images taken in 2015 suffered from bad A/D converters, which introduced additional errors. Furthermore, the exposure time calculator underestimated the exposure time and thus gave a lower imaging depth. A lot of updates were made during the past two years, including the telescope control, data acquisition software, ancillary observing tools, remote monitoring, etc. The observation progress is shown in Figure 5 of \citet{zou17}. The fractions of the completed observations for each pass and each filter are given in Table \ref{tab-obspar}. We have completed about 41\% of the whole survey, including 47\% of the $g$-band observations and 35\% of the $r$-band observations.

\subsection{Image quality}
\begin{figure*}[!ht]
\begin{center}
\includegraphics[width=0.8\textwidth]{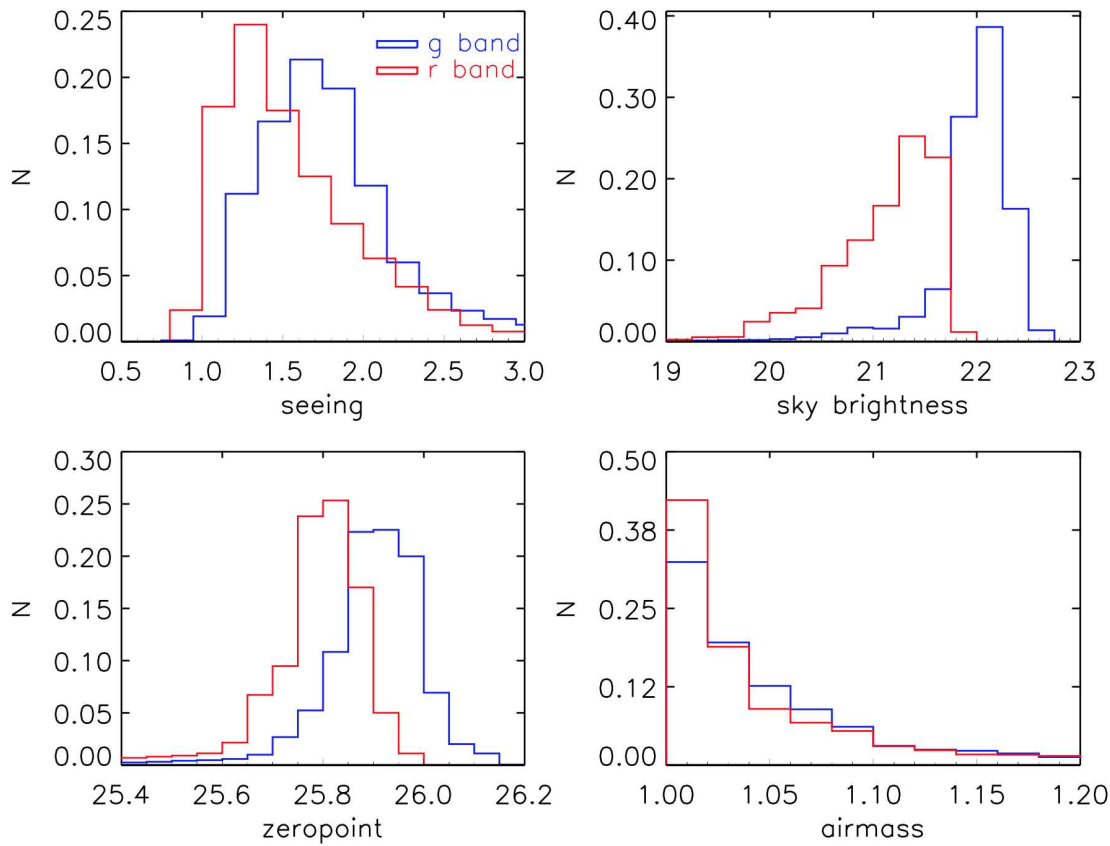}
\caption{\boldtext{Normalized distributions of seeing (FWHM in arcsec), sky background (in mag arcsec$^{-2}$), zeropoint (in mag), and airmass for $g$ (blue) and $r$ (red) band observations.} }\label{fig-obspar}
\label{image_quality}
\end{center}
\end{figure*}

The observations are controlled by the observing strategy and ancillary observing tools as described earlier \citep[see also in][]{zou17}. Figure \ref{fig-obspar} shows the distributions of the observational parameters for the two bands, including seeing, sky brightness, zeropoint, and airmass. The median  values of these parameters for each pass are also provided in Table \ref{tab-obspar}. The median seeing is about 1\arcsec.7 for $g$ and 1\arcsec.5 for $r$. The $r$-band seeing is typically better. The median $g$-band sky brightness is 22.0 mag arcsec$^{-2}$, and the median $r$-band sky brightness is about 21.2 mag arcsec$^{-2}$. The $g$-band observations were usually taken in dark time, and most of the $r$-band observations were taken in grey time. The airmass is close to 1.0 and the median zeropoints in these two bands are 25.9 and 25.8 mag for 1 e/s. These parameters are used to estimate the depths of single-epoch images, and subsequently determine which tiles to be re-observed in future runs. In Table \ref{tab-obspar}, the median single-epoch depths (5$\sigma$) estimated by the photometric errors are also listed, which will be analyzed in Section \ref{sec-analyses}.
\begin{table*}[!ht]
\caption{Observational parameters (median values) for different passes and filters} \label{tab-obspar}
\begin{tabular}{c|cccc|cccc}
\hline
\hline
Filter & \multicolumn{4}{c|}{$g$} & \multicolumn{4}{c}{$r$} \\ \cline{1-9} 
\multirow{2}{*}{\backslashbox{Parameter}{Pass} }    & Pass 1 & Pass 2 &  Pass 3 & All & Pass 1 & Pass 2&  Pass 3 & All \\
           & 62\% & 58\% &  21\% & 47\% & 47\% &  46\% &  12\% & 35\% \\
\hline
seeing (FWHM in arcsec)       &   1.55 &  1.83   &   1.90 &   1.71 &   1.30 &   1.61 &   1.52 &   1.45 \\
sky (mag arcsec$^{-2}$)           & 22.05 & 22.06  & 21.97 & 22.04 & 21.39 & 21.15 & 21.02 & 21.24 \\
airmass     &   1.05 &  1.04   &   1.02 &   1.04 &   1.03 &   1.03 &   1.01 &  1.03 \\
zeropoint (mag)   & 25.90 & 25.91  & 25.93 & 25.91 & 25.82 & 25.79 & 25.72 & 25.80 \\
depth (mag)        & 23.38 & 23.33 & 23.31 & 23.35 & 22.97 & 22.89  & 22.80 & 22.92 \\
\hline
\end{tabular}
\tablenotetext{}{\boldtext{The BASS footprint is covered by three passes. Each pass is taken under different observational conditions: Pass 1 for both photometric nights and seeing $< 1\arcsec.7$; Pass 2 for either photometric nights or seeing $> 1\arcsec.7$; Pass 3 for any other cases. Here ``All" means that the values are derived from all single-epoch images regardless of passes. }}
\end{table*}

\section{Data reduction} \label{sec-reduction}
\subsection{Image Processing}
\begin{figure*}[ht]
\centering
\includegraphics[width=\textwidth]{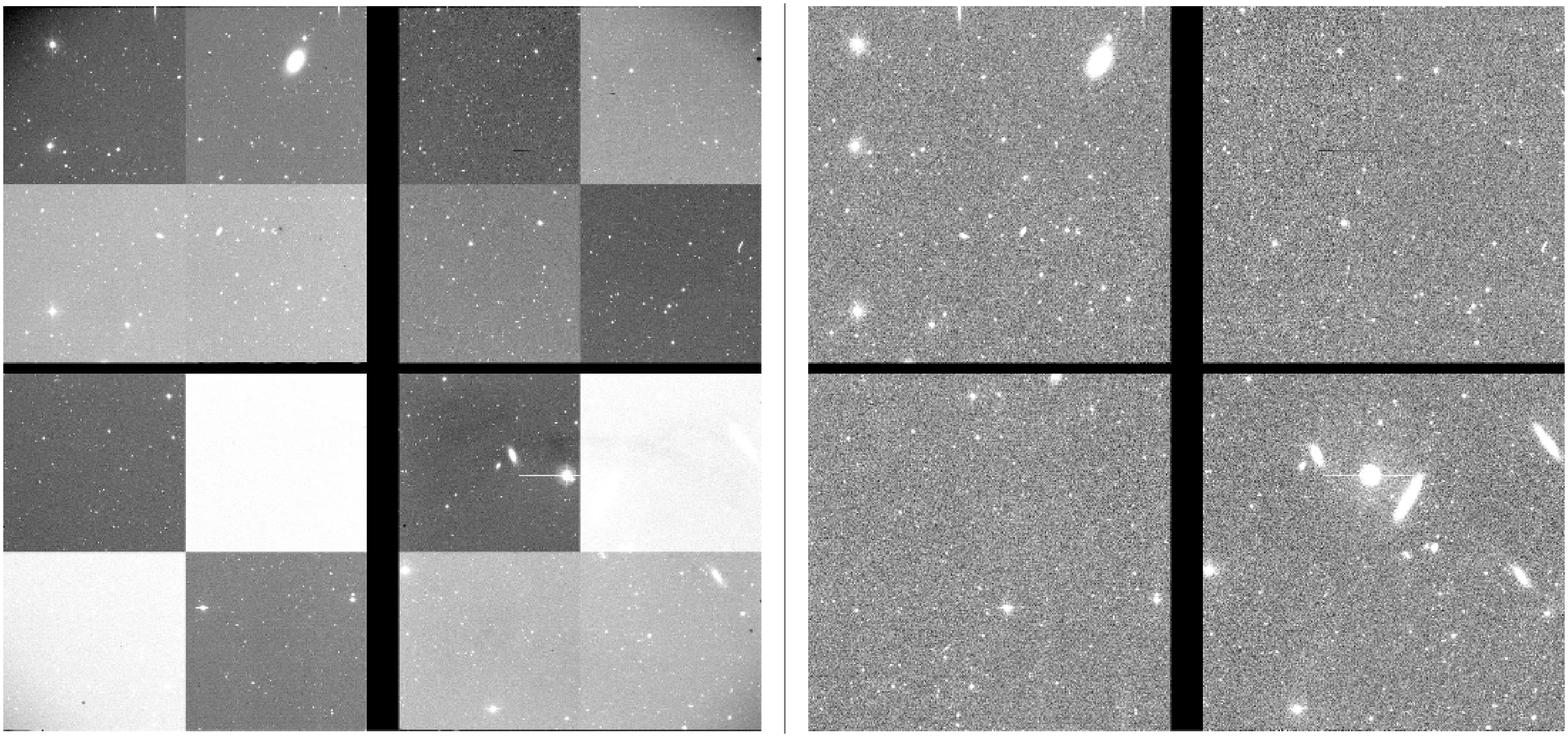}
\caption{An example of the raw (left) and processed (right) science images.} \label{fig-detrend}
\end{figure*}

In this section, we present some special handling and updates relative to the SCUSS image processing described in \citet{zou15}. Figure \ref{fig-detrend} illustrates one raw science image and its fully processed image. In addition to the calibrated science images, the weight and flag maps (denoting the masking pixels) are also provided. The units of the image counts is e/s.

\subsubsection{Overscan, bias and flat corrections}
\begin{figure}[!ht]
\centering
\includegraphics[angle=0,width=\columnwidth]{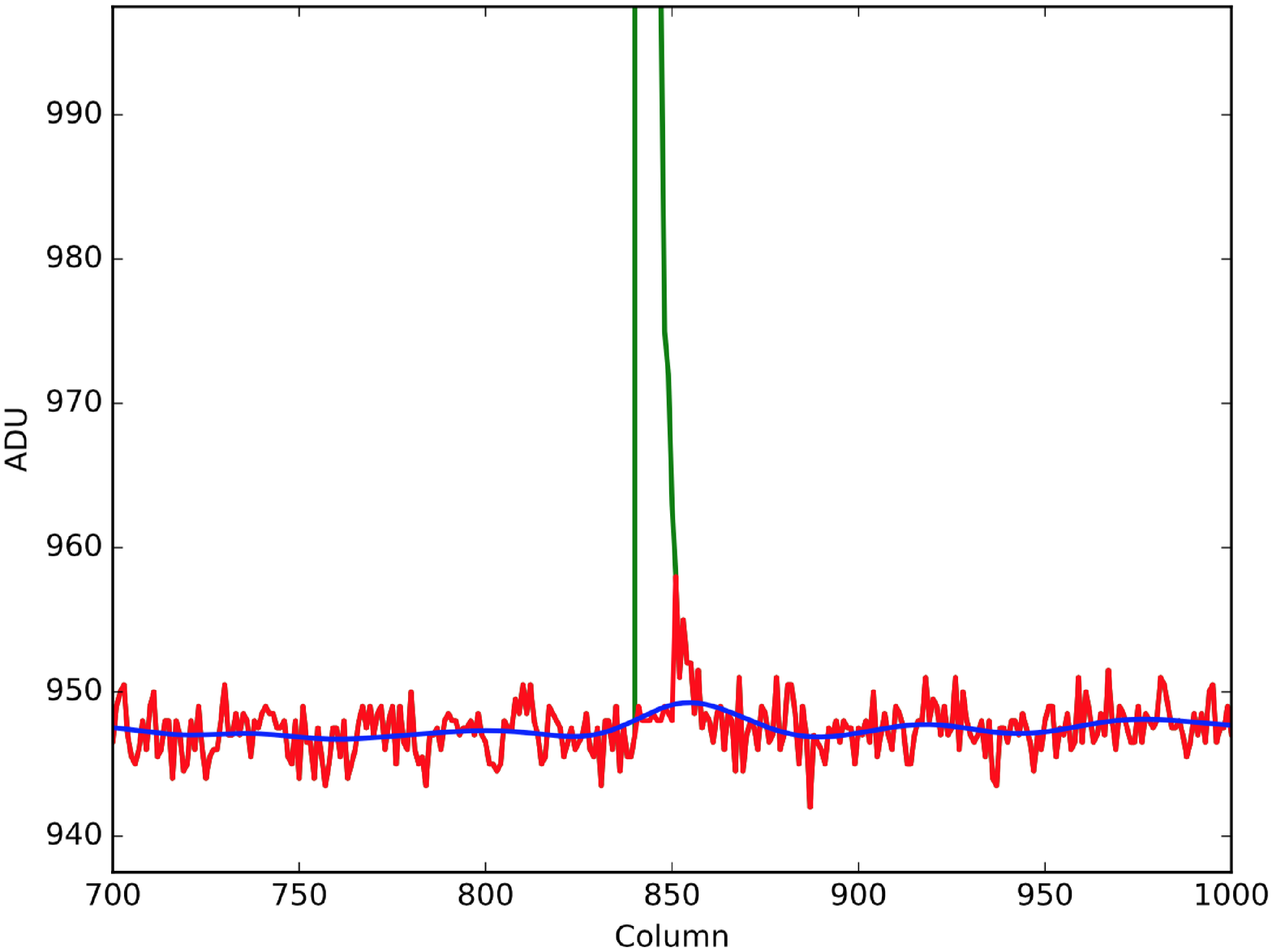}
\caption{An example for the X-direction overscan in one BASS raw image contaminated by bright stars. The green line is the median overscan line and the red one is the one after masking and interpolating the region contaminated by a bright star (showing an abrupt transition). The blue line gives the smoothed overscan.} \label{fig-overscan}
\end{figure}

Comparing with the data taken in 2015, the BASS images in 2016 have two overscan regions in both CCD X and Y directions. The X-direction overscan is usually contaminated by bright stars (see the green line in Figure \ref{fig-overscan}). The contaminated regions are masked and interpolated, and then the  overscan line is smoothed as shown in blue in Figure \ref{fig-overscan}. A two-dimensional overscan image is created by averaging the smoothed X and Y-direction overscans. The median bias frame is generated with about 20 bias images taken before and after the night observations. Different from the SCUSS data reduction, we  combine the dome flat and night sky flat to do flat-fielding. The dome flats have high signal-to-noise (S/N) ratios, which are used to correct the pixel-to-pixel sensitivity variations. The night sky flat is obtained by calculating the median of all science images taken during the night. It is regarded to have the closest optical path to science images, but with a lower S/N. This night sky flat is used for the illumination correction. 

\subsubsection{Crosstalk}
\begin{figure}[!ht]
\centering
\includegraphics[angle=0,width=\columnwidth]{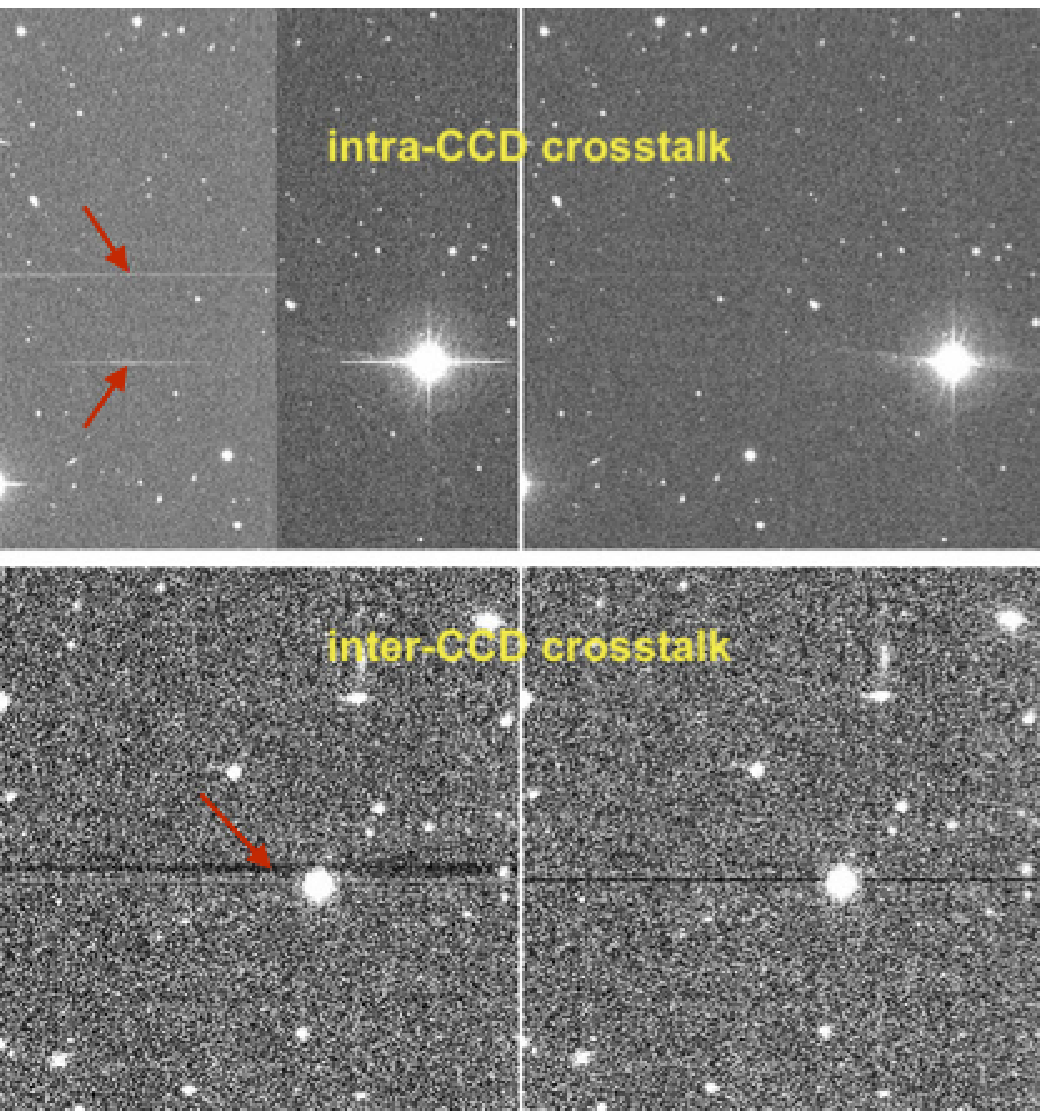}
\caption{The intra-CCD (top) and inter-CCD (bottom) crosstalk effects in the BASS images. The left images in top and bottom panels present the crosstalk signals shown by the red arrows, and the right images show the results after corrections. }\label{fig-crosstalk}
\end{figure}

The 90Prime detector consists of four CCDs in an array of $2\times2$, which are nearly centrosymmetric. Each CCD has four amplifiers at the CCD corners for reading out. The crosstalk effect can be easily seen as mirror images of saturated stars. We find that there are both intra-CCD and inter-CCD crosstalk effects, showing positive or negative signals in images. We use a large number of images to estimate the crosstalk coefficients. First, bright stars are detected in one amplifier. The crosstalk signals are then identified in all other amplifiers, and the crosstalk coefficients are calculated by minimizing the difference between the signals and sky background. The overall level of the crosstalks is 5:10000. The average level of the intra-CCD and inter-CCD crosstalks are 18:10000 and 2:10000, respectively. The crosstalk correction in a particular amplifier is computed by summing the pixel values in all other amplifiers multiplied by their crosstalk coefficients and then subtracted from this amplifier. Table \ref{tab-crosstalk} lists the crosstalk coefficients for all pairs of amplifiers. Figure \ref{fig-crosstalk} show examples of the crosstalk effect and corresponding corrections. 

\begin{table*}[!ht]
\caption{Crosstalk coefficients for all pairs of amplifiers.} \label{tab-crosstalk}
{\small
  \begin{tabular}{ccccccccccccccccc}
  \tableline\tableline
HDU & 1 & 2 & 3 & 4 & 5 & 6 & 7 & 8 & 9 & 10 & 11 & 12 & 13 & 14 & 15 & 16\\
\tableline
1 & 0 & -23 & -33 & -30 & 1 & 2 & 2 & 2 & 1 & -1 & 1 & 1 & 2 & 2 & 1 & 3\\
2 & -17 & 0 & -24 &-23 &1 & 1 &2 & 3 & 1 & 0 &1 & 1 & 2 & 3 & 3 & 3\\ 
3 & -10 & -8 & 0& -11& 1 & 1& 2 & 2 & 1 & 0 &1 &1 &2 &2 &2 &4\\ 
4 & -11 & -9 & -11 & 0 &2 & -4 &2 & 2 & 1 & 0 & 0 & 1 & 1 & 2 & 2 & 2\\ 
5 & 3& 3&3&3&0&-33&-37&-27&2&-1&2&1&0&0&1&2\\ 
6 & 3&3&4&3&-10&0&-17&-14&3&2&4&3&1&1&1&2\\ 
7 &3&3&3&3&-11&-14&0&-6&2&1&2&2&1&2&2&2\\ 
8 & 3&3&4&4&-9&-6&-5&0&1&3&4&2&0&1&1&1\\ 
9 & 3&2&3&2&1&1&2&2&0&-16&-23&-19&3&3 &2&2\\ 
10 & 2&1&2&1&0&1&1&1&-31&0&-40&-32&2&2&2&2\\ 
11 & 3&2&2&1&0&-1&1&1&-25&-16&0&-17&2&2&1&2\\ 
12 & 3&2&3&2&1&1&2&2&-15&-11&-8&0&-2&1&2&1\\ 
13 &4 &3&4&3&-1&0&1&1&-9&0&1&3&0&-24&-26&-22\\ 
14 &4&3&4&3&0&1&1&2&-9&1&1&-2&-24&0&-26&-22\\ 
15 &4&4&5&4&1&1&1&2&-11&0&1&3&-13&-10&0&-5\\ 
16&  5&5&5&5&-1&1&1&2&-7&-3&-1&0&-15&-14&-12&0\\ 
  \tableline
  \end{tabular}
  \tablenotetext{}{The scale is 10$^{-5}$.}}
\end{table*}

\subsubsection{CCD artefacts}
\begin{figure}[!ht]
\centering
\includegraphics[angle=0,width=\columnwidth]{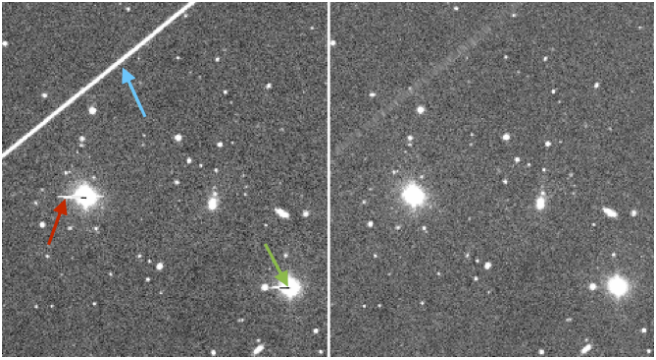}
\caption{Artefact identification and interpolation. The left and right panels show the same part of an image before and after artefact corrections. The green arrow points to a black core of a saturated star, and the red arrow shows a bleeding trail from another saturated star. The cyan arrow designates a satellite track.} \label{fig-artefacts}
\end{figure}

The CCD artefacts in the BASS images include bad pixels, cosmic rays, satellite trails, the black cores and bleeding trails of saturated stars. Bad pixels are determined using a series of dome flats with increasing exposure time. A linear fit to the pixel values as a function of exposure time is performed. The pixels with non-linearity are considered as bad pixels. Cosmic rays are identified by their different sharpness from astronomical sources. The bleeding trails of saturated stars are easily detected due to their ADU values close to the CCD saturation level. The satellite trails are identified as straight lines by using the Hough transform. We find that five amplifiers (HDU 5, 6, 7, and 8 in CCD \#2 and HDU 15 in CCD \#4) have a problem of integer overflow during A/D converting. The core of saturated stars show low warped counts. All artefacts are masked and recorded in a flag image (see Section \ref{sec-access}) and relevant pixels are interpolated. Figure \ref{fig-artefacts} gives an example of masking black cores, bleeding trails, and satellite tracks. 

\subsection{Astrometry}
\subsubsection{Astrometry by SCAMP}
The Software for Calibrating AstroMetry and Photometry \citep[SCAMP;][]{ber06} is utilized to derive astrometric solutions for the BASS images. SCAMP reads the catalogs generated by SExtractor \citep{ber96} and calculates the astrometric solution by matching detected objects to reference catalogs. The SDSS catalog contains many more faint sources than other catalogs such as the Forth US Naval Observatory CCD Astrograph Catalog \citep[UCAC4]{zac13}, and most of the BASS area is covered SDSS, so the SDSS catalog is used as the main reference catalog. If a region is out of the SDSS coverage, the point-source catalog of the Two Micron All Sky Survey \citep[2MASS;][]{skr06} is used as the reference. These regions are usually close to the Galactic plane, so there are  plenty of stars for accurate measurements of astrometric solutions. We adopt the ``TAN" projection and a 4th-order polynomial distortion. The distortion in BASS images varies with time as the telescope focus gradually changes due to the telescope gravitational deformation. When the telescope slews a large distance, the focus and corresponding focal distortion change significantly. Therefore, The astrometric solution is independently calculated for each BASS image by SCAMP.

\subsubsection{Astrometric accuracy} 
\begin{figure}[!ht]
    \includegraphics[width=\columnwidth]{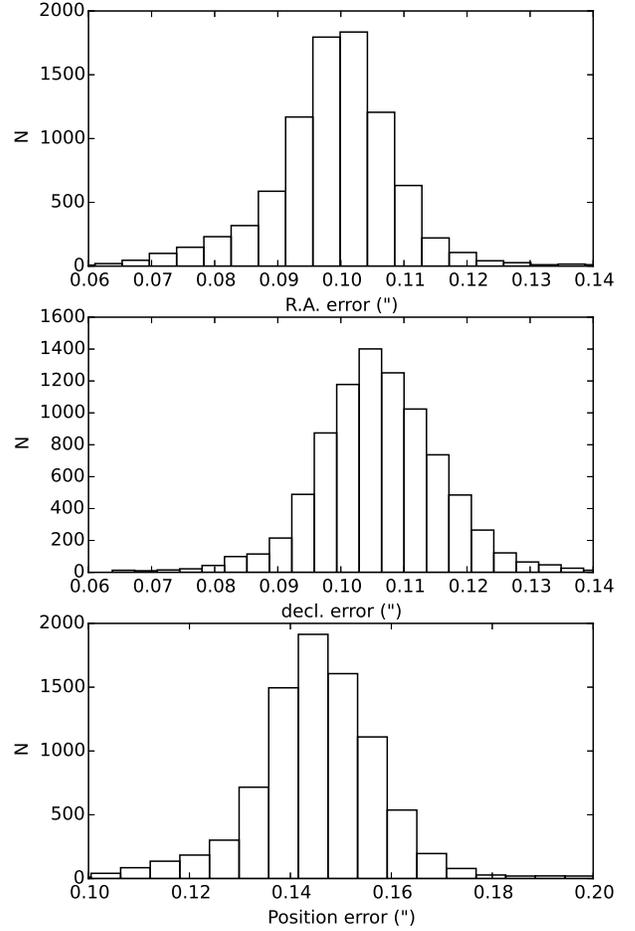} 
    \caption{The distributions of astrometric errors in arcsec. The top, middle, and bottom panels show the astrometric errors in R. A. and decl. and the global positional error, respectively. }\label{fig-astromerror}
\end{figure}

The main factors that affect the accuracy of the astrometric solution include: (a) pattern matching between input and reference catalogs; (b) the accuracy of the source positional measurements; (c) the coordinate accuracy in reference catalogs; (d) the stability of the focal distortion. For each BASS image, we estimate the astrometric errors by comparing the coordinates of detected objects with those in the reference catalog. There are average 1,300 stars ($r$ down to $\sim21$ mag) for each image and a total of $\sim$ 9,000 images to obtain the distributions of astrometric errors as shown in Figure \ref{fig-astromerror}. The median errors in R. A. and decl. are about 0\arcsec.10 and 0\arcsec.11, respectively. The median positional error as presented in the bottom panel of Figure \ref{fig-astromerror} is about 0\arcsec.15. We plan to use more accurate astrometric data from Gaia as the reference catalog \citep{gai16a} in future data releases. Gaia had its first data release in September 2016 \citep{gai16b} with a source magnitude limit of 20.7 mag. In addition, we also plan to fix the focal distortion for the images that are taken in a certain period of time. In this way, a large number of stars can be used to derive a more accurate distortion map. 

\subsubsection{CTE effect on astrometry}  \label{sec-cte}
\begin{figure*}[!ht]
\centering
	\subfigure{\includegraphics[width=0.45\textwidth]{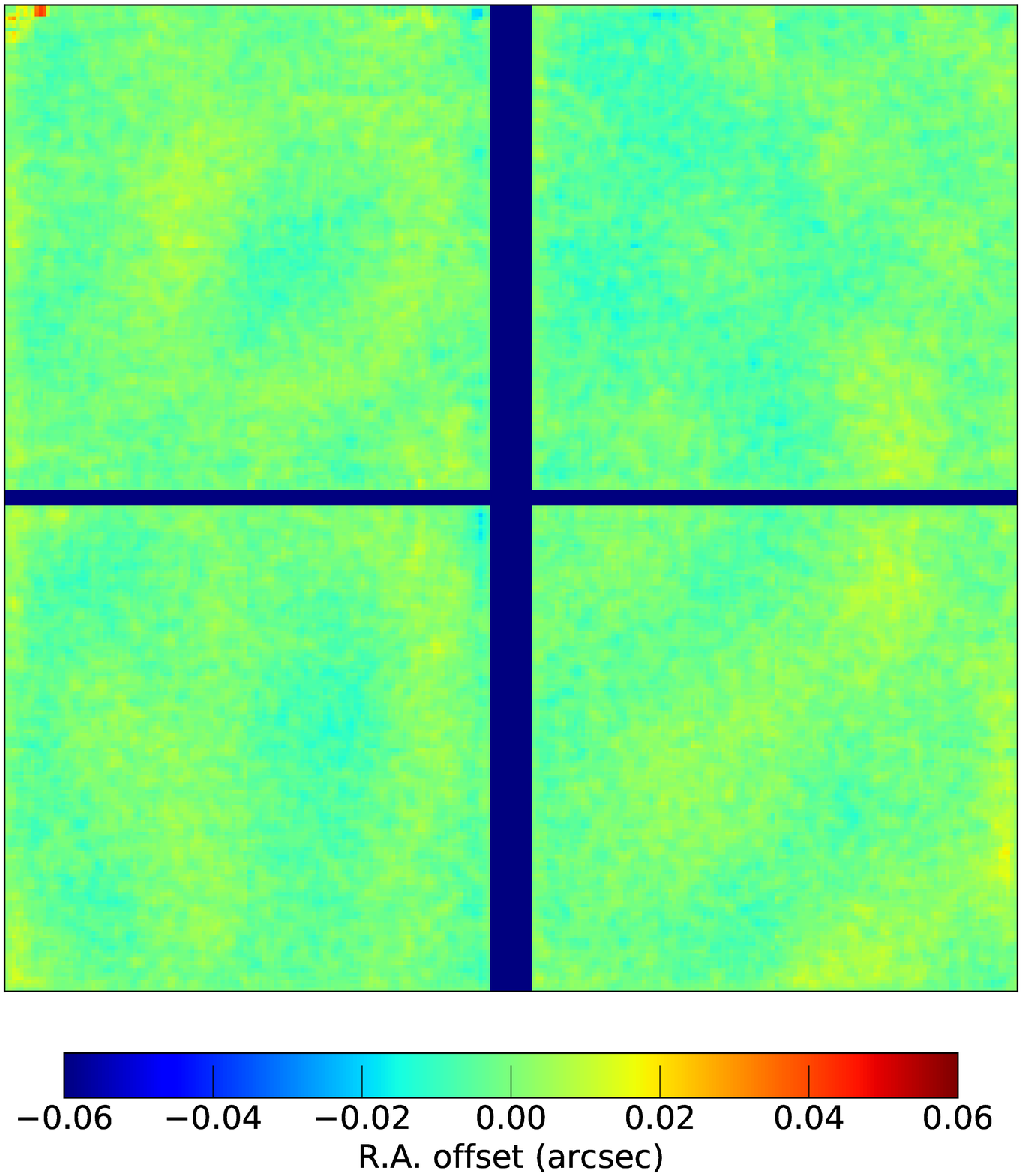}}
	\subfigure{\includegraphics[width=0.45\textwidth]{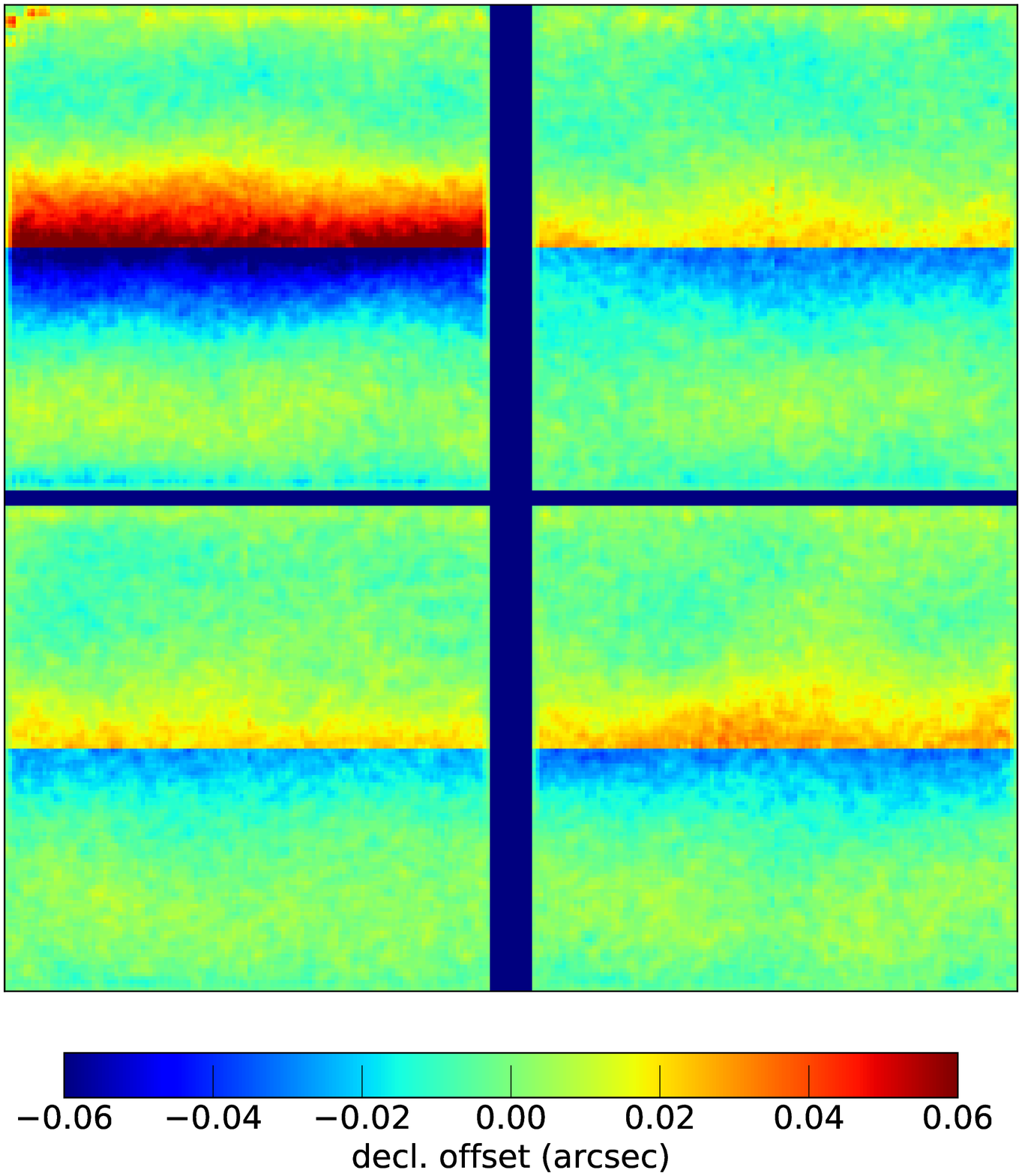}}
	\caption{Astrometric residual maps in R. A. (left) and decl. (right) relative to the reference catalog. The color bar below gives the astrometric offset ranging from -0\arcsec.06 to 0\arcsec.06. } \label{fig-cteeffect}
\end{figure*}

The charge transfer efficiency (CTE) is believed to have impact on astrometric accuracy and should be corrected in the catalogs. During the reading out, each CCD line is transferred to the previous lines in parallel, and a small fraction of charges are left over there. The remaining charges will be transferred together with the charges of the next line. In this way, the charges received for each line keep a part of remaining charges from all previous lines, so that the centroid of sources deviate towards the direction of reading out. We have made some simulations with sources of Gaussian profiles that are evenly distributed over the image. The mock charges are transferred as the way described above. We find that the centroids of sources gradually deviate away from the initial positions, and the positional offsets reach the maximum for the last line. For CTE $=$ 0.99998, the largest offset is about 0.03 pixel after transferring 2000 lines. The CTE effect can be clearly seen in the BASS astrometric residual maps. This is also reported in \citet{mun14}. Figure \ref{fig-cteeffect} shows the residual maps in R. A. and decl. derived by matching objects to the SDSS catalogs. More than ten thousands of images and their relatively bright and isolated stars are used to obtain these maps. The Y axis (north is up) is the reading-out direction. The CCD transfers charges with four amplifiers from both upper and lower sides, so the centroid shifts of the top and bottom parts are opposite. Thus, it can be seen that there is a big declination offset in the middle of the images, while there is no obvious substructures in R.A. The CCD \#1 suffers most seriously from the CTE effect. We use these maps to correct the source coordinates in the photometric catalogs. 

\subsection{Photometric calibration and accuracy}
The BASS photometric solutions are tied to Pan-STARRS1 (hereafter PS1), which has released its first data \citep{cha16}. A large aperture of 13-pixel radius ($\sim 5\arcsec.9$) adapting to the BASS seeing is adopted to calculate photometric zeropoints. We use SExtractor to detect sources with S/N $>$ 5 and internal flags $<2$, and measure their aperture magnitudes by subtracting local sky background. The sources are cross-matched with the point-source catalogs of PS1. The cross-matching radius is 2\arcsec. The PS1 stars are required to be detected at least once in all filters and the median PSF magnitudes are used. 

We also obtain the transformation equations to transform the PS1 magnitudes to the BASS photometric system. First, zeropoints are estimated without any systematic transformation. After applying zeropoints, the magnitude difference as a function of the $g - i$ color is derived. We use a third-order polynomial fit to determine the color terms in the range of $0.4 < (g_{\rm PS1}-i_{\rm PS1}) < 2.7$, which are presented in Equation (1)--(3). 
\begin{figure*}[!ht]
\begin{eqnarray}
    (g-i) = g_{\rm PS1}-i_{\rm PS1}, \\
    g_{\rm BASS} = g_{\rm PS1} - 0.08826 + 0.10575(g-i) - 0.02543(g-i)^2 + 0.00226(g-i)^3,  \\
    r_{\rm BASS} = r_{\rm PS1} + 0.07371 - 0.07650(g-i) + 0.02809(g-i)^2 - 0.00967(g-i)^3.
\end{eqnarray}
\end{figure*}
Figure \ref{fig-colorterm} shows the magnitude difference between the BASS and PS1 as a function of $g - i$. The polynomial fitting to the data is also over-plotted in this figure. These color terms are applied to all CCDs.
\begin{figure}[!ht]
	\centering
	\includegraphics[width=\columnwidth]{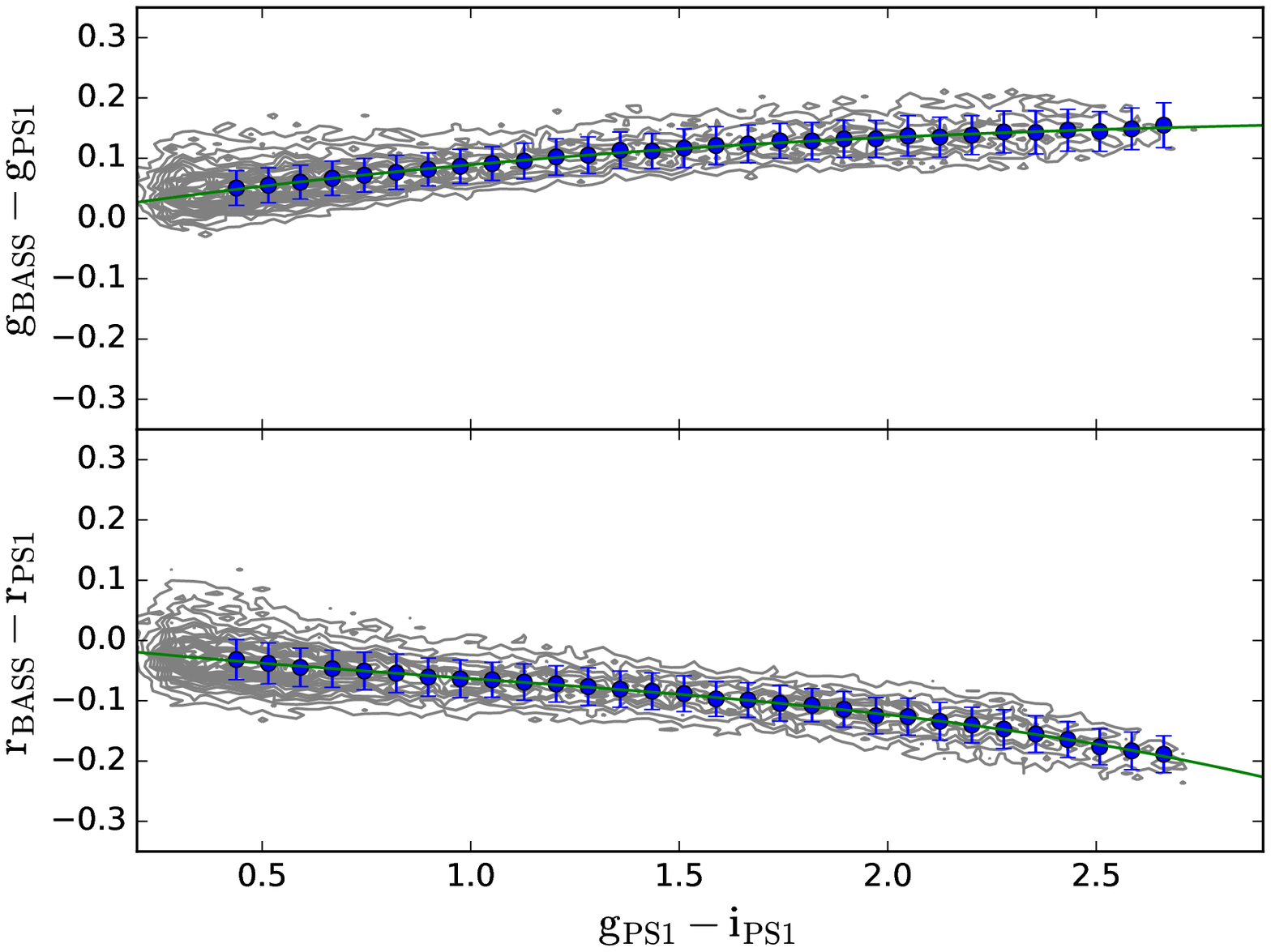}
	\caption{\boldtext{The magnitude differences between BASS and PS1 as a functions of $g - i$.} The upper panel is for the BASS $g$ band, while the lower panel is for the $r$ band. The contours show the densities of data points. The blue points with error bars give the median magnitude differences and the standard deviations in the magnitude bins. A third-order polynomial fitting to these points is also shown in the green solid curve. } \label{fig-colorterm}
\end{figure}

\begin{figure*}[!ht]
	\centering
	\includegraphics[width=0.45\textwidth]{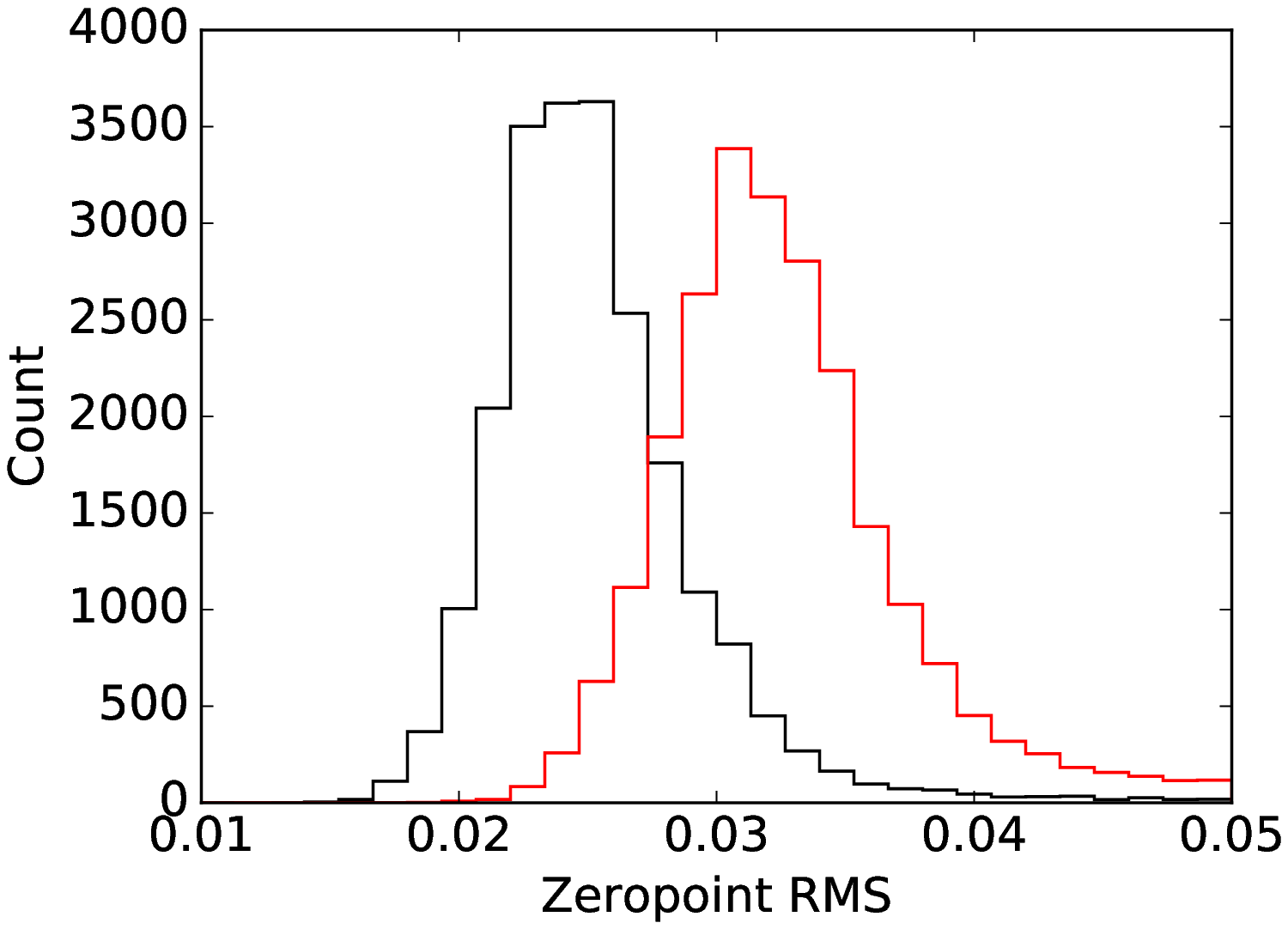}
	\includegraphics[width=0.45\textwidth]{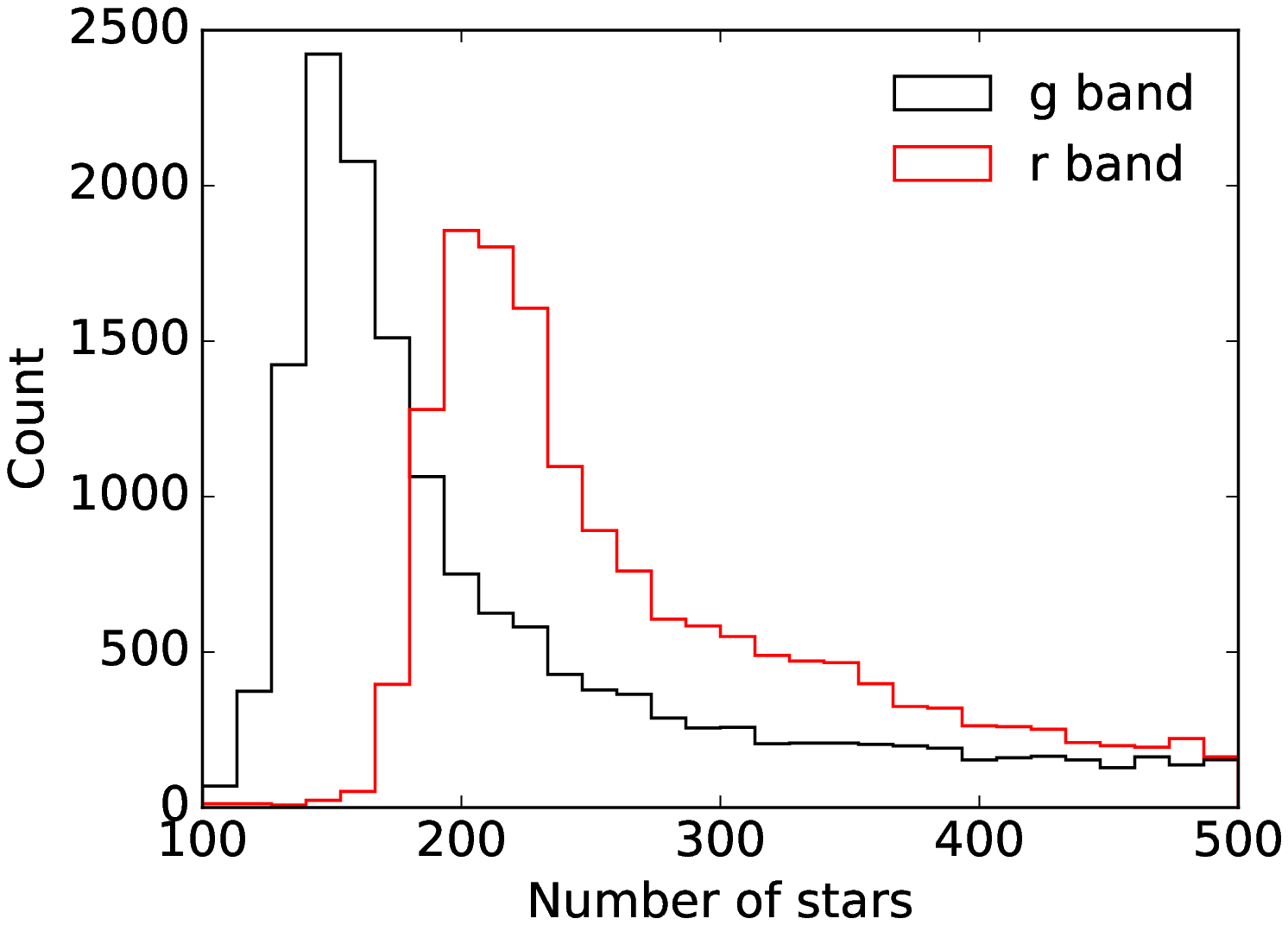}
	\caption{Left: Distribution of the zeropoint (in mag) RMS. Right: number distribution of the stars that are matched to calculate the zeropoints.}
	\label{fig-zperror}
\end{figure*}

The median zeropoint for each CCD image is calculated using a $3\sigma$-clipping algorithm. Figure~\ref{fig-zperror} presents the histograms of the numbers of stars used to calculate the zeropoints and the zeropoint RMS. In general, in each image there are 120--250 stars for the $g$ band and 180--300 stars for the $r$ band that are matched to PS1. The median RMS values of the zeropoints for the $g$ and $r$ bands are about 0.025 mag and 0.032 mag, respectively. In addition to the external calibration using reference catalogs (e.g. SDSS and PS1), BASS has three overlap exposures for each field, and some observations on photometric nights for randomly selected fields over the BASS footprint. The photometric zeropoint accuracy is expected to be better than 1\% after the internal calibration is applied. The method of the internal photometric calibration will be implemented in the future data reduction.

\section{Photometry} \label{sec-photometry}
The photometry in this data release is performed on single-exposure images, which includes fixed aperture magnitudes, automatic aperture magnitudes from SExtractor, and PSF magnitudes. Since the observing strategy makes different passes of the same area that have distinct image qualities (especially seeing), we cannot obtain accurate photometry from stacked images. The photometry on single-exposure images is more favourable: the PSF profile across an image is relatively simple to model, and there is no interpolation that may introduce additional error in the process of resampling and coaddition. The magnitudes measured in single images are co-added to make object catalogs with full depths. 

\subsection{Photometry on single-exposure images}
\subsubsection{SExtractor photometry} \label{sec-sex}
SExtractor is used for source detections and aperture magnitude measurements in single-exposure images. Table \ref{tab-sexpar} presents some key configuration parameters for SExtractor. In order to detect as many faint objects as possible, the detection threshold is 1$\sigma$ above the sky background and the default Gaussian kernel is used for image convolution. The minimum contrast parameter for deblending is set to be 0.0005 and the cleaning efficiency is set to be 0.5 to clear some fake sources around bright stars and very extending objects. The aperture radii for fixed aperture photometry are almost same as used in SCUSS. There are 12 apertures with radii of about 1\arcsec.36, 1\arcsec.82, 2\arcsec.27, 2\arcsec.72, 3\arcsec.63, 4\arcsec.54, 5\arcsec.90, 7\arcsec.92, 9\arcsec.08, 11\arcsec.35, 13\arcsec.62, and 18\arcsec.26. The effective gain is equal to the exposure time as the units of the calibrated images is e/s, which will directly affect the error estimation of the aperture magnitudes. Weight and flag images are fed as inputs to SExtractor. The magnitudes are aperture corrected with light growth curves from bright stars as described in \citet{zou15}.

\begin{table*}[!ht]
\centering
\caption{Key parameters as SExtractor inputs \label{tab-sexpar}}
\begin{tabular}{lll}
\hline
\hline
Parameter & Value & Meaning \\
\hline
DETECT\_MINAREA   & 3  & Minimum number of pixels above the threshold\\
DETECT\_THRESH & 1.0 & Detection threshold\\
FILTER\_NAME & gauss\_2.0\_3x3.conv & Image convolution kernel\\
DEBLEND\_NTHRESH & 64 & Number of deblending sub-thresholds \\
DEBLEND\_MINCONT & 0.0005 & Minimum contrast parameter for deblending \\
CLEAN\_PARAM & 0.5 & Cleaning efficiency \\
WEIGHT\_TYPE  &  MAP\_WEIGHT & Type of weight \\
PHOT\_APERTURES   & 6,8,10,12,16,20,26,34.884,40,50,60,80   & aperture diameters in pixels \\
BACK\_SIZE  &  256 & Size of the background mesh grid \\
\hline
\end{tabular}
\end{table*}

\subsubsection{PSF photometry}
A new code for the PSF fitting is specially developed, which is based on the source positions from SExtractor and PSF models from PSFEx \citep{ber06}. The Sky background and its RMS map are calculated by using the same algorithm as the SExtractor background estimating method. The BASS images are convolved with a Gaussian kernel ($\sigma = 1.5$ pixels).  Signal pixels above 1$\sigma$ of the background are detected. The watershed segmentation algorithm is adopted to separate objects that are close to each other. Starting from specific markers, this algorithm treats pixel values as a local topography and flood basins from the markers until basins attributed to different markers meet on watershed lines \footnote{\url{http://scikit-image.org/docs/dev/auto_examples/segmentation/plot_watershed.html}}. Here the markers are SExtractor detected source positions. The PSF fitting will treat objects differently depending on whether they are isolated (i.e. no pixel connecting to others). If an object is isolated, it will be directly fitted with the PSF model, otherwise objects connected to each other will be fitted simultaneously. These objects are ranked according to the brightness, and the three brightest objects are fitted first. 

\subsubsection{Photometric residual maps} \label{sec-photres}
\begin{figure}[!ht]
\includegraphics[width=\columnwidth]{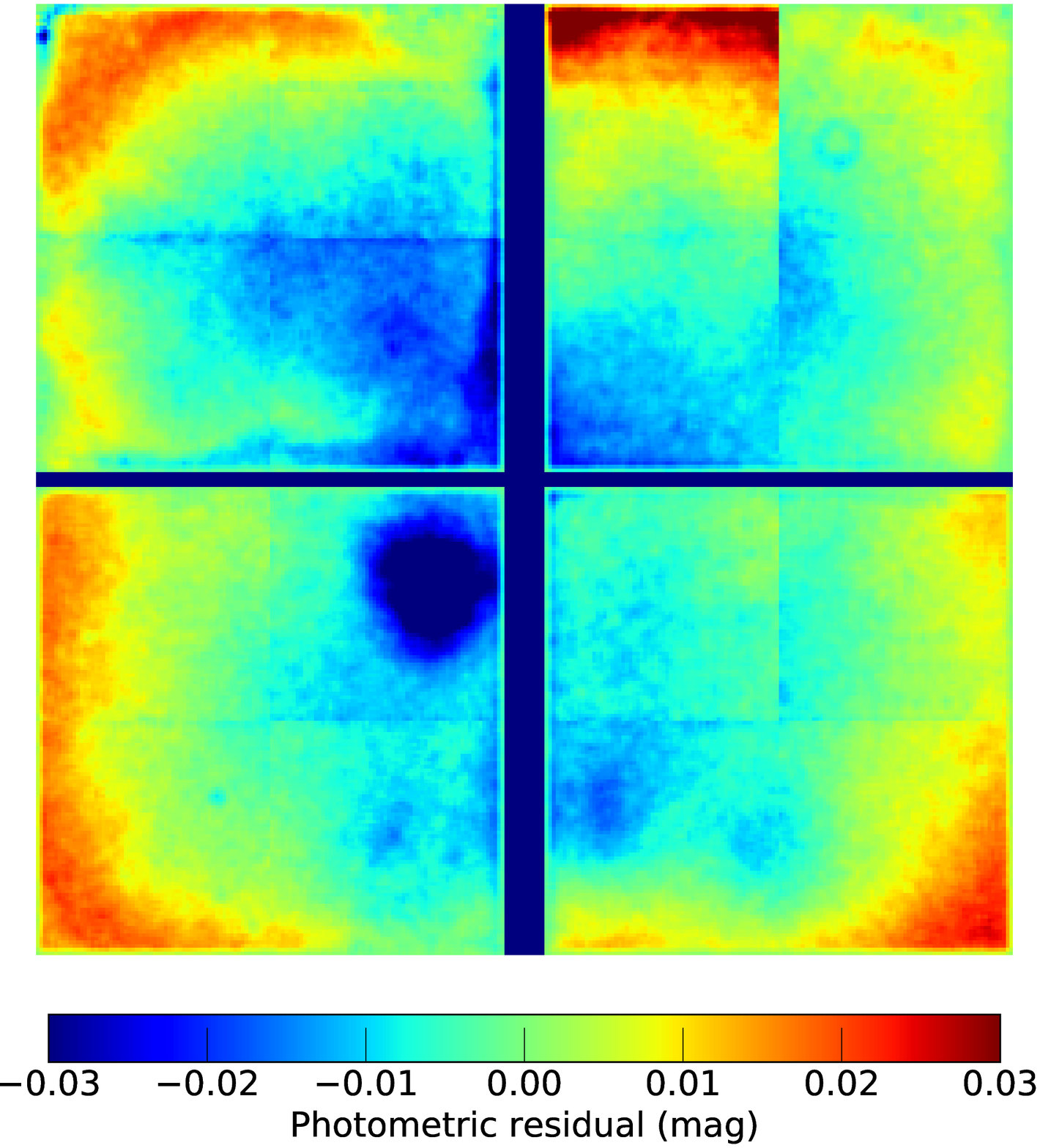}
\caption{Photometric residual maps for the $g$ band. The color bar gives the magnitude range from --0.03 to 0.03 mag. } \label{fig-photres}
\end{figure}

We construct the photometric residual maps for each band and each CCD by comparing the BASS PSF magnitudes with the PS1 catalogs. Millions of bright (but not saturated) isolated stars in more than 10,000 images are used. These photometric residual structures are mainly caused by the focal plane distortion, which are approximately centrosymmetric in the camera FoV. In addition, the imperfect flat-fielding, scattered light, and the CTE effect may also contribute a little to the residual maps. Figure \ref{fig-photres} shows the photometric residuals for the BASS $g$ band. The maps for the $r$ band are similar. We use these maps to correct all magnitudes in the catalogs of single-epoch images.

\subsection{Co-added catalogs} \label{sec-coadd}
We divide the BASS footprint into evenly distributed blocks. Each block has a size of 0\arcdeg.68$\times$0\arcdeg.68 (5400$\times$5400 pixels with a pixel scale of 0\arcsec.454) and an overlap of 0\arcdeg.02 in both right ascension and declination directions. The detailed process of co-adding single-epoch catalogs related to the blocks is described below. 

(a) For a given block, all related single-epoch catalogs are collected according to the qualities of corresponding images: (1) seeing $< 4\arcsec.0$; (2) external astrometric errors in R.A. and decl. are less than 0\arcsec.5; (3) numbers of stars used for the zeropoint calculation are larger than 100; (4) the estimated depths are fainter than 20 mag. 

(b) Objects near the edges of the CCD images (within 20 pixels) are removed. The BASS tiles are designed to be overlapped with more than 30\arcsec, so the CCD edges are also covered by more than three exposures and the depths there would still meet the requirement.

(c) All single-epoch catalogs related to the block are cross-matched with a matching radius of 1.0\arcsec.  Objects are regarded as real sources if $n_\textrm{match} >= 2$ or if  $n_\textrm{match} = 1$ and $n_\textrm{exposure} = 1$. Here $n_\textrm{match}$ is the total number of matched objects regardless of filters, and $n_\textrm{exposure}$ is the number of exposures at that place. The coordinates of the matched objects are averaged to form a unique source.

(d) The magnitude and shape measurements for a unique source from different single-epoch catalogs are weight-averaged to generate the co-added measurements. The magnitude co-addition and the estimate of the corresponding weighted error are performed in the linear flux space. We also calculated their standard error, which is the parameter RMS divided by the square root of the number of objects. More detailed content about the single-epoch and co-added catalogs can be found on the public webpage of the data release \footnote{\url{http://batc.bao.ac.cn/BASS/doku.php?id=datarelease:home}}

\section{Analyses} \label{sec-analyses}
\subsection{Single-epoch depths}
\begin{figure}[!ht]
\begin{center}
\includegraphics[width=\columnwidth]{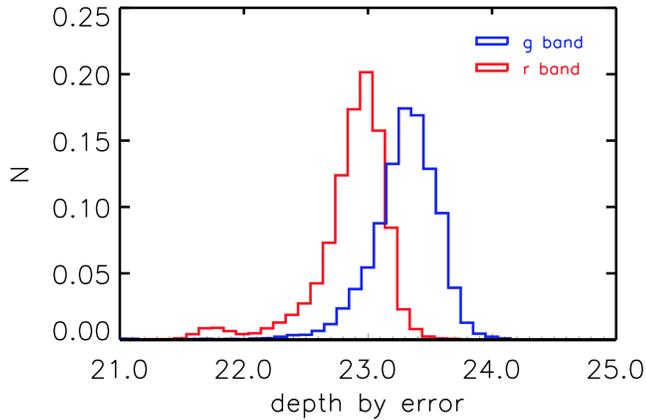}
\caption{The histograms of single-epoch depths in the $g$ (blue) and $r$ (red) bands estimated by the PSF magnitude error at S/N = 5.}\label{fig-singdepth}
\end{center}
\end{figure}

The magnitude limits for single-epoch images are estimated by the PSF magnitude error at about 0.21 mag, which corresponds to the S/N of 5. Figure \ref{fig-singdepth} displays the distributions of these magnitude limits. The median $g$ and $r$-band depths are 23.35 and 22.92 mag, respectively. The depths for each pass and filter are also shown in Table \ref{tab-obspar}. The full depths are expected to be 0.6--0.7 mag deeper. All depths presented in this paper are corrected for the Galactic extinction, if not specified. We use the extinction coefficients of $A_g = 3.303$ and $A_r=2.285$ and the reddening map from \citet{sch98}. It should be noted that it is hard to separate point sources from extended ones with faint magnitudes, so the depths might be underestimated for point sources but overestimated for extended ones. 

\begin{figure}[!ht]
\begin{center}
\includegraphics[width=\columnwidth]{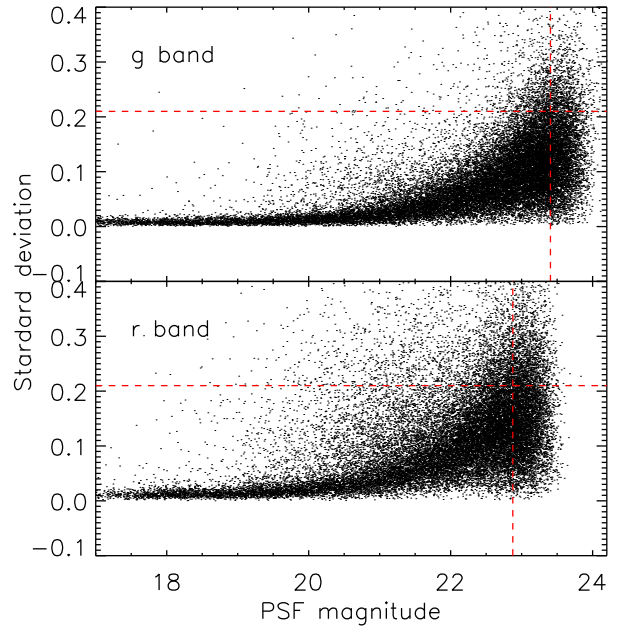}
\caption{The magnitude standard deviation as a function the PSF magnitude for objects observed three times (top for $g$ band and bottom for $r$ band). The horizontal line gives the deviation of 0.21 mag, and the vertical line shows the corresponding magnitude limits. }\label{fig-depthstd}
\end{center}
\end{figure}

The BASS footprint is imaged in three passes, so each object can be observed at least three times. The standard deviation of magnitudes for the same object gives an actual measurement of the photometric error. Figure \ref{fig-depthstd} shows the standard deviation as a function of the PSF magnitude. The magnitude limits estimated as the median magnitudes at the deviation of 0.21 mag (S/N = 5) are 23.41 mag for $g$ and 22.87 mag for $r$. They are approximately consistent with the depths estimated by the PSF magnitude error. 

\subsection{Full depths}
\begin{figure}[!ht]
\begin{center}
\includegraphics[width=\columnwidth]{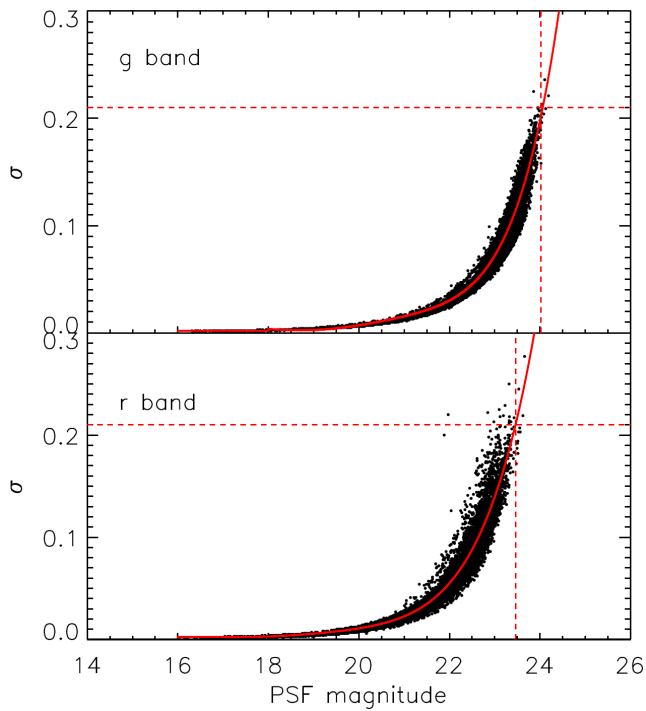}
\caption{Photometric errors of objects with three exposures as functions of the PSF magnitudes for both BASS $g$ (top) and $r$ (bottom) bands. The horizontal dashed line shows the error of 0.21 mag, which corresponds to the S/N of 5, while the vertical line gives the corresponding full depth estimation. The solid curves polynomial fits to the data.\label{fig-fulldepth}}
\end{center}
\end{figure}

We estimate the $g$ and $r$-band full depths from the co-added catalogs in the area covered by three passes with typical observational conditions. The depth is calculated with the photometric error of the co-added PSF magnitude, which is derived by three repeated measurements. Figure \ref{fig-fulldepth} shows the magnitude errors as a functions of magnitudes for the two bands. The full depths at S/N = 5 are 24.06 (g) and 23.46 (r). There are a lack of sources close to these limits, mainly because we detect objects in the single-epoch images and generate the co-added catalog by cross-matching. In the future, we will detect objects in stacked images and measure the magnitudes in single-epoch images with the prior information of positions. The depth values are also dependent on which photometric system the data related to. Our depths are tied to the PS1 system, which has systematic magnitude offsets of about 0.14 and 0.03 mag for $g$ and $r$ bands relative to the SDSS.
 
\subsection[]{Comparison with the SDSS and DECaLS}
\begin{figure}[!ht]
\begin{center}
\includegraphics[width={1.0\columnwidth}]{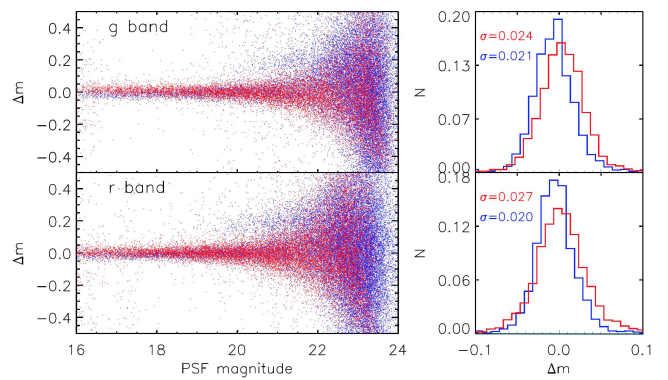}
\caption{The magnitude differences as a function of the BASS PSF magnitude in the $g$ (upper) and $r$ (lower) bands. The points in blue present the comparison between  BASS and DECaLS, and the red points show the comparison between  BASS and SDSS.  The histograms show the normalized magnitude difference distributions with the PSF magnitude between 17 and 20 mag.} \label{fig-magcomp}
\end{center}
\end{figure}

BASS is 1--2 magnitude deeper than the SDSS, but is slightly shallower than DECaLS. The data from these two surveys can be used for comparison of the photometric accuracy and depth with BASS. We select the region located in the NOAO Deep Wide-Field Survey field \citep{jan04}, which has been covered by the three surveys. The point sources are accurately modelled by PSF profiles, and the PSF magnitudes from the SDSS and BASS are used for analyses. The DECaLS magnitudes are derived from the model fluxes in the DR3 catalogs \footnote{\url{http://legacysurvey.org/dr3/}}. The system transformations have been applied to all magnitudes. Figure\ref{fig-magcomp} shows the magnitude comparisons between  BASS and other two surveys. The scatters (1$\sigma$ RMS) of objects with magnitude between 18 and 21 are about 0.03 for both $g$ and $r$ bands, although the scatter between BASS and DECaLS is somewhat smaller. 

\begin{figure*}[!ht]
\includegraphics[width=1.0\textwidth]{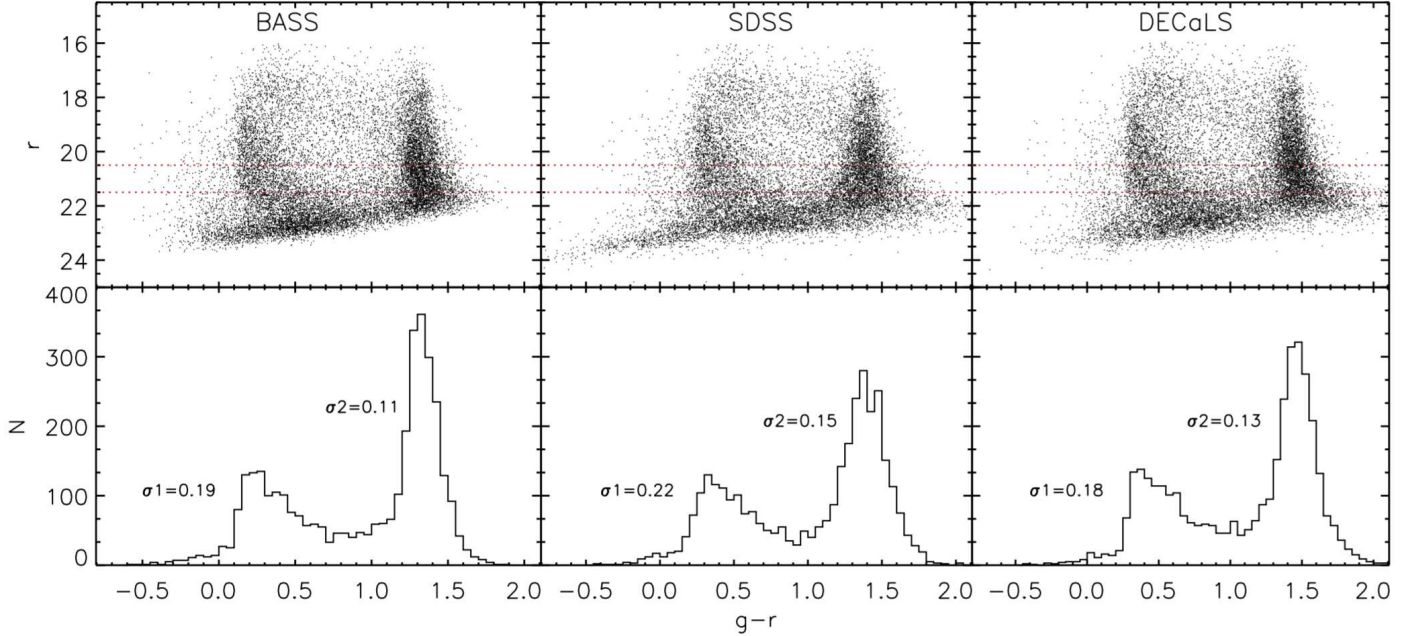}
\caption{Color magnitude diagrams for the BASS (left), SDSS (middle), DECaLS (right) in the upper panels. Same matched point sources are shown in these plots. The bottom panels shows the $g-r$ distributions for the three survey data at $r \sim 21$ mag ($20.5 < r < 21.5$), where $\sigma_1$ and $\sigma_2$ are fitted Gaussian standard deviations of the two peaks. }
\label{fig-cmd}
\end{figure*}

The color--magnitude diagram (CMD) is often used for studying the Galactic structures. We also use the CMD to compare the photometry of point sources from the three surveys. Figure \ref{fig-cmd} shows the CMDs of $g - r$ vs. $r$ for BASS, SDSS and DECaLS. We select point sources with the SDSS star/galaxy separation as well as the star-like type in the BASS  co-added catalogs. The BASS type are derived as the average of the SExtractor ``CLASS\_STAR" parameters measured in the single-epoch images (type value $>$ 0.8). \boldtext{The bottom panels of Figure \ref{fig-cmd} show the $g - r$ distributions for these three surveys at $r \sim 21$ mag. Two populations of Galactic stars can be clearly seen as two peaks. We calculate scatters of the two peaks as standard deviations of fitted Gaussian functions. From these plots and scatters,  we can see that the SDSS is shallower than BASS and DECaLS and the BASS and DECaLS data both performs similarly well in their CMDs.} 
 
{\bfseries 
\section{Data Products and Data Access} \label{sec-access}
\subsection{Single-epoch images}
The single-epoch images in our data products are detrended science frames with astrometric and photometric solutions stored in the FITS header. Each FITS file has only one CCD image without any extension, which has the pixel number of 4096$\times$4032 (about $0\arcdeg.52\times0\arcdeg.51$) and pixel scale of 0$\arcsec$.454. The count unit for each pixel is DN/s. The filename is something like ``p5482g0100\_1.fits", starting with a letter ``p" following with last 4 digits of the Julian date, the filter name, 4 digits of the run number from the CCD data acquiring software, and the CCD number connected with a underscore. The filter name can be either ``g" or ``r". In the image, north is up and east is left. For the astrometry, we adopt the ``TAN" projection with a fourth-order polynomial distortion (a series of PV keywords in the FITS header). The photometric zeropoint is given by a keyword ``CCDZPT" and its RMS error relative to PS1 is presented by a keyword ``CCDPHRMS". The number of stars used for determining the zeropoint is provided by a keyword of ``NMATCH" . The magnitude of an object can be calculated as $m = -2.5 \mathrm{log}_{10} F + \mathrm{CCDZPT}$, where $F$ is the total data counts. The observation conditions such as airmass, seeing, and sky brightness are also provided in the FITS header. We provide some key information of each single-exposure images in a binary FITS table\footnote{\url{http://batc.bao.ac.cn/BASS/lib/exe/fetch.php?media=datarelease:dr1:bass-dr1-ccdinfo.fits}} and the corresponding column description can be found in the webpage\footnote{\url{http://batc.bao.ac.cn/BASS/doku.php?id=datarelease:dr1:dr1_ccdinfo:home}}. The images with the column ``IMQ" = 1 is used to form the co-added catalogs as described in Section \ref{sec-coadd}. Other images have either bad calibrations or shallow depths. 

Corresponding to each CCD image, there are two other images: a weight (named as  ``p5482g0100\_1.wht.fits") and flag maps (named as ``p5482g0100\_1\_od.fits"). They have the same size as the single-epoch image. The weight map gives the pixel-level inverse variance, consisting of the readout-noise, poisson noise, and error propagation during detrending. The flag image tags unreliable pixels, which are coded in decimal as a sum of powers of 2: 
\begin{itemize}
\item 1 bad pixels
\item 2 saturated pixels or pixels located in bleeding trails of saturated stars
\item 4 pixels contaminated by cosmic rays
\item 8 pixels contaminated by satellite tracks
\end{itemize}

\subsection{Catalogs}
The single-exposure catalogs contain the mixing photometric information from SExtractor and our PSF photometric code.  The magnitudes are derived by the automatic aperture, fixed aperture and PSF photometry. The shape parameters and morphological classification come from SExtractor. There are 12 fixed circular apertures as presented in Section \ref{sec-sex}. The equatorial  coordinates in J2000 and all magnitudes are respectively corrected with astrometric and photometric residuals as described in Sections \ref{sec-cte} and \ref{sec-photres}. The detailed description of the catalog columns is shown in the webpage\footnote{\url{http://batc.bao.ac.cn/BASS/doku.php?id=datarelease:dr1:singe_column:home}}. 

The co-added catalogs are generated by cross-matching of the single-epoch catalogs and co-adding all sorts of measurements. The naming of a catalog file is related to the center of a block defined as a box in equatorial coordinates, which has a size of about $0\arcdeg.68\times0\arcdeg.68$ (see Section \ref{sec-coadd}). Taking ``182.932+36.060\_75841-gradd.fits" as an example, the first 7 characters give the R.A. with an accuracy to three decimal places, the following 7 characters give the decl. with a sign, and the 5 digits connected with a underscore present the block ID. We can find the block information (e.g. coordinates of the center and four corners) from a binary FITS table\footnote{\url{http://batc.bao.ac.cn/BASS/lib/exe/fetch.php?media=datarelease:dr1:bass-dr1-blocks-prior.fits}} and its column description in the webpage\footnote{\url{http://batc.bao.ac.cn/BASS/doku.php?id=datarelease:dr1:dr1_blockinfo:home}}.  The co-added catalogs contain photometric and positional contents for both $g$ and $r$. The Galactic extinction from \citet{sch98} for each object is included.  We provide both the weighted errors and standard errors for all photometric magnitudes. In addition, the number of exposures at the specified position and the number of valid measurements that are co-added are also given in the catalogs. The description of the catalog columns are presented in the webpage \footnote{\url{http://batc.bao.ac.cn/BASS/doku.php?id=datarelease:dr1:coadd_column:home}}. One should note that there are quite a few missing faint objects close to the magnitude limits because the sources are detected only on single-exposure images. We will detect sources in the stacked images in future data releases. Besides, the blocks have overlapped of about 1\arcmin.2, so objects are not unique although the object IDs in the catalogs are unique. 

\subsection{Data Access}
All BASS data products and survey information can be retrieved on our public data release website\footnote{\url{http://batc.bao.ac.cn/BASS/doku.php?id=datarelease:home}}. The access service is implemented by the Chinese Astronomical Data Center\footnote{\url{http://explore.china-vo.org/}}. There are two ways to get images or catalogs: one is to search data with form-based queries and the other is to download data through direct links into the directory trees. The following list presents the access links: 
\begin{itemize}
\item Form-based queries
   \begin{itemize}
   \item single-epoch images: \url{http://explore.china-vo.org/data/bassdr1images/f}
   \item single-epoch catalogs: \url{http://explore.china-vo.org/data/bassdr1photo/f}
   \item co-added catalogs: \url{http://explore.china-vo.org/data/bassdr1gradd/f}
   \end{itemize}
\item Directory trees
   \begin{itemize}
   \item single-epoch images: \url{http://casdc.china-vo.org/archive/BASS/DR1/single-epoch-images/}
   \item single-epoch catalogs: \url{http://casdc.china-vo.org/archive/BASS/DR1/single-epoch-catalogs/}
   \item co-added catalogs: \url{http://casdc.china-vo.org/archive/BASS/DR1/coadded-catalogs/}
   \item bulk download: \url{http://batc.bao.ac.cn/BASS/doku.php?id=datarelease:dr1:dr1_wgetbulk:home}
   \end{itemize}
\end{itemize}
The single-epoch images and catalogs in the directory trees are distributed in a series of subdirectories named as the local date when the data were taken. We can generate a ``wget" script in the link of bulk download, where all related data in a box of the equatorial coordinates will be found. 
}

\section{Summary} \label{sec-summary}
BASS is a new imaging survey in the north Galactic cap, conducted by the NAOC and Steward Observatory. The survey uses the 2.3 m Bok telescope on Kitt Peak to cover an area of about 5400 deg$^2$ in the SDSS $g$ and DES $r$ bands. The expected depths (5$\sigma$) corrected for the Galactic extinction are 24.0 and 23.4 mag, respectively. The observations started in January 2015 and have undergone for two years. In total, about 41\% of the whole survey has been completed. This paper describes the data reduction and the first data release.  

Compared to the data reduction pipelines by SCUSS, which used the same telescope and camera to survey the south Galactic cap in the $u$ band and released the data in December of 2015 \citep{zou15,zou16}, BASS pipelines for DR1 include some updates and different photometric methods:
\begin{itemize}
\item Two-dimensional overscan images are constructed and subtracted. The combination of dome flats and super sky flats is used for flat-fielding. Both intra-CCD and inter-CCD crosstalk coefficients are derived. The intra-CCD crosstalk effect is significantly larger than the inter-CCD one. CCD artefacts are masked and interpolated, including  bad pixels, cosmic rays, satellite trails, low counts at the center of saturate stars, and bleeding trails from those stars. Weight and flag images are also generated. 
\item The astrometric solutions are derived by SCAMP with the 4th-order polynomial distortions. The distortion is independently determined for each image. The reference catalogs are the SDSS or 2MASS. The overall position error is about 0\arcsec.15.
\item The photometric solutions are tied to point source catalogs from PS1. The median zeropoint RMS is about 0.03 for the both $g$ and $r$ bands. 
\item The photometry is performed on single-epoch images. Both fixed and automatic aperture magnitudes are derived by SExtractor. A new PSF fitting code is developed to measure PSF magnitudes based on the object positions detected by SExtractor. The co-added catalogs are generated by cross-matching the photometric catalogs and merging the measurements. This photometric mode is better than the photometry performed on stacked images, because the BASS observing strategy leads to considerably different image qualities for three passes. 
\end{itemize}

The median $5\sigma$ magnitude limits of single-epoch images are about 23.4 and 22.9 mag for the $g$ and $r$ bands, respectively. The full depths are 24.1 and 23.5 mag for the two bands. The current mode of photometry and co-adding results in a lack of faint sources with magnitudes close to the full depths. The data can be accessed via the public webpage \footnote{\url{http://batc.bao.ac.cn/BASS/doku.php?id=datarelease:dr1:home}}. For the future data releases, we will detect sources in stacked images and co-add photometric measurements derived in single-epoch images with prior information from those sources. In addition, we will include updates on the astrometry and zeropoint calculations.

\acknowledgments

The BASS is a collaborative program between the National Astronomical Observatories of Chinese Academy of Science and Steward Observatory of the University of Arizona. It is a key project of the Telescope Access Program (TAP), which has been funded by the National Astronomical Observatories of China, the Chinese Academy of Sciences (the Strategic Priority Research Program ``The Emergence of Cosmological Structures" Grant No. XDB09000000), and the Special Fund for Astronomy from the Ministry of Finance. The BASS is also supported by the External Cooperation Program of Chinese Academy of Sciences (Grant No. 114A11KYSB20160057) and Chinese National Natural Science Foundation (Grant No. 11433005). The BASS data release is based on the Chinese Virtual Observatory (China-VO).

This work is also supported by the Chinese National Natural Science Foundation grant Nos. 11203031, 11373033, 11333003, 11390372, 11433005, 11673027, and 11603034, and by the National Basic Research Program of China (973 Program), Nos. 2014CB845704, 2014CB845702, and 2013CB834902.

SDSS-III is managed by the Astrophysical Research Consortium for the Participating Institutions of the SDSS-III Collaboration including the University of Arizona, the Brazilian Participation Group, Brookhaven National Laboratory, Carnegie Mellon University, University of Florida, the French Participation Group, the German Participation Group, Harvard University, the Instituto de Astrofisica de Canarias, the Michigan State/Notre Dame/JINA Participation Group, Johns Hopkins University, Lawrence Berkeley National Laboratory, Max Planck Institute for Astrophysics, Max Planck Institute for Extraterrestrial Physics, New Mexico State University, New York University, Ohio State University, Pennsylvania State University, University of Portsmouth, Princeton University, the Spanish Participation Group, University of Tokyo, University of Utah, Vanderbilt University, University of Virginia, University of Washington, and Yale University.

The Pan-STARRS1 Surveys (PS1) and the PS1 public science archive have been made possible through contributions by the Institute for Astronomy, the University of Hawaii, the Pan-STARRS Project Office, the Max-Planck Society and its participating institutes, the Max Planck Institute for Astronomy, Heidelberg and the Max Planck Institute for Extraterrestrial Physics, Garching, The Johns Hopkins University, Durham University, the University of Edinburgh, the Queen's University Belfast, the Harvard-Smithsonian Center for Astrophysics, the Las Cumbres Observatory Global Telescope Network Incorporated, the National Central University of Taiwan, the Space Telescope Science Institute, the National Aeronautics and Space Administration under Grant No. NNX08AR22G issued through the Planetary Science Division of the NASA Science Mission Directorate, the National Science Foundation Grant No. AST-1238877, the University of Maryland, Eotvos Lorand University (ELTE), the Los Alamos National Laboratory, and the Gordon and Betty Moore Foundation.

This publication makes use of data products from the Two Micron All Sky Survey, which is a joint project of the University of Massachusetts and the Infrared Processing and Analysis Center/California Institute of Technology, funded by the National Aeronautics and Space Administration and the National Science Foundation.

\end{document}